\documentclass[%
  a4paper, twocolumn, prl, superscriptaddress, amsmath, amssymb, floatfix
]{revtex4-1}
\usepackage{graphicx}
\usepackage{epstopdf}
\usepackage[table]{xcolor}
\usepackage{braket}
\usepackage[colorlinks,bookmarksopen=false]{hyperref}
\usepackage[caption=false]{subfig}
\usepackage{bbm}
\usepackage{makecell}
\usepackage{lipsum, blindtext}
\usepackage{xr}
\usepackage[normalem]{ulem}
\hypersetup{%
  pdfstartview={FitW},
  linkcolor=blue,
  citecolor=blue,
  filecolor=black,
  menucolor=black,
  urlcolor =black
}

\newcommand{\im}{\ensuremath{\textup{i}}}

\newcommand{\rp}{\ensuremath{\mathfrak{Re}}}
\newcommand{\ip}{\ensuremath{\mathfrak{Im}}}

\newcommand{\op}[1]{\ensuremath{\mathsf{#1}}}

\newcommand{\uop}{\op{\mathbbm{1}}}

\newcommand{\pulse}{\mathcal{E}}
\newcommand{\dmap}{\mathcal{D}}
\newcommand{\Nc}{N_{\mathrm{c}}}

\newcommand{\rZ}{\op{\rho}(0)}
\newcommand{\rS}{\op{\rho}_{\mathrm{S}}}
\newcommand{\rB}{\op{\rho}_{\mathrm{B}}}

\newcommand{\OS}{\op{O}_{\mathrm{S}}}
\newcommand{\OB}{\op{O}_{\mathrm{B}}}
\newcommand{\Oc}{\op{O}_{\mathrm{c}}}

\newcommand{\xx}{\op{\sigma}_{1} \otimes \op{\sigma}_{1}}
\newcommand{\xy}{\op{\sigma}_{1} \otimes \op{\sigma}_{2}}

\newcommand{\yx}{\op{\sigma}_{2} \otimes \op{\sigma}_{1}}
\newcommand{\yy}{\op{\sigma}_{2} \otimes \op{\sigma}_{2}}

\newcommand{\zx}{\op{\sigma}_{3} \otimes \op{\sigma}_{1}}
\newcommand{\zy}{\op{\sigma}_{3} \otimes \op{\sigma}_{2}}

\newcommand{\tr}{\ensuremath{\mathrm{tr}}}

\newcommand{\change}[1]{\textcolor{black}{#1}}

\begin{document}

\title{%
  Fundamental Bounds on Qubit Reset
}

\author{Daniel Basilewitsch}
\affiliation{%
  Theoretische Physik, Universit\"{a}t Kassel, D-34132 Kassel, Germany
}
\affiliation{%
  Dahlem Center for Complex Quantum Systems and Fachbereich Physik, Freie
  Universit\"{a}t Berlin, Arnimallee 14, D-14195 Berlin, Germany
}

\author{Jonas Fischer}
\affiliation{%
  Theoretische Physik, Universit\"{a}t Kassel, D-34132 Kassel, Germany
}
\affiliation{%
  Dahlem Center for Complex Quantum Systems and Fachbereich Physik, Freie
  Universit\"{a}t Berlin, Arnimallee 14, D-14195 Berlin, Germany
}
\affiliation{%
  Laboratoire Interdisciplinaire Carnot de Bourgogne (ICB), Universit\'e de
  Bourgogne-Franche Comt\'e, F-21078 Dijon Cedex, France
}

\author{Daniel M. Reich}
\affiliation{%
  Theoretische Physik, Universit\"{a}t Kassel, D-34132 Kassel, Germany
}
\affiliation{%
  Dahlem Center for Complex Quantum Systems and Fachbereich Physik, Freie
  Universit\"{a}t Berlin, Arnimallee 14, D-14195 Berlin, Germany
}

\author{Dominique Sugny}
\affiliation{%
  Laboratoire Interdisciplinaire Carnot de Bourgogne (ICB), Universit\'e de
  Bourgogne-Franche Comt\'e, F-21078 Dijon Cedex, France
}

\author{Christiane P. Koch}
\email{christiane.koch@fu-berlin.de}
\affiliation{%
  Theoretische Physik, Universit\"{a}t Kassel, D-34132 Kassel, Germany
}
\affiliation{%
  Dahlem Center for Complex Quantum Systems and Fachbereich Physik, Freie
  Universit\"{a}t Berlin, Arnimallee 14, D-14195 Berlin, Germany
}

\date{\today}

\begin{abstract}
  Qubit reset is a basic prerequisite for operating quantum devices, requiring
  the export of entropy. The fastest and most accurate way to reset a qubit is
  obtained by coupling the qubit to an ancilla on demand. Here, we derive
  fundamental bounds on qubit reset in terms of maximum fidelity and minimum
  time, assuming control over the qubit and no control over the ancilla. Using
  the Cartan decomposition of the Lie algebra of qubit plus \change{two-level}
  ancilla, we identify the types of interaction and controls for which the qubit
  can be purified. For these \change{configurations}, we show that
  a time-optimal protocol \change{consists of purity exchange between qubit and
  ancilla brought into resonance}, where the maximum fidelity is identical for
  all cases but the minimum time depends on the type of interaction and control.
  \change{Furthermore, we find the maximally achievable fidelity to increase
  with the size of the ancilla Hilbert space, whereas the reset time remains
  constant.}
\end{abstract}

\maketitle

The ability to initialize qubits from an arbitrary mixed state to a fiducial
pure state is a basic building block in quantum information (QI)
science~\cite{divincenzo:2000}. Initializing the qubit, or - equivalently
- resetting it after completion of a computational task, requires some means to
export entropy. At the same time, for device operation, the qubit needs to be
well-protected and isolated from its environment. It is thus not an option to
simply let the qubit equilibrate with its environment; rather, active reset is
indispensable. A common approach to actively initialize a qubit uses projective
measurements~\cite{Nielsen}, but for many QI architectures, this suffers from
being slow, see e.g. Refs.~\cite{riste:2012,johnson:2012} for the example of
superconducting qubits. Rapid reset is made possible by coupling each qubit to
an ancilla in a tunable way. This can be a fast decaying state, such as in laser
cooling, or an auxiliary system such as another qubit~\cite{Boykin:2002,
basilewitsch:2017} or a resonator~\cite{geerlings:2013, magnard:2018,
Egger2018}. Then, the coupling strength together with either the switching time
for the coupling or the ancilla decay rate determine the overall time required
to reset the qubit. This bound on the reset protocol duration is a specific
instance of open quantum system speed limits~\cite{reviewQSL}. At present, the
overall time required for qubit reset presents \change{one of the main
limitations for device operation, in particular for superconducting qubits}. On
the other hand, in these architectures, there exists a great flexibility in the
design of tunable couplings between qubit and ancilla~\cite{KjaergaardARCMP20}.
This raises the question which type of coupling allows for the most accurate and
fastest reset.

Here, we answer this question, \change{assuming no external control over the
ancilla. Starting with a two-level ancilla, where} the reset dynamics is an
element of SU(4), we leverage earlier results on the quantum optimal control in
SU(4)~\cite{khaneja:2001, romano:2006, romano:2007, romano:2007b}. Making use of
the Cartan decomposition of SU(4)~\cite{alessandrobook, gilmore}, we identify
the qubit-ancilla couplings \change{which allow for qubit purification.} For all
Hamiltonians fulfilling this criterion, we use quantum optimal control
theory~\cite{Glaser:15} to determine the controls on the qubit that realize
a time-optimal reset. \change{Moreover, we identify the dynamics for maximum
purity reset when the ancilla Hilbert space dimension is larger than two.
\change{Our combination of geometric, algebraic and numerical tools} provides
a new perspective on the problem of qubit reset, identifying ultimate
performance limits for duration and accuracy.}

\change{Our focus on using an ancilla to reset the qubit is motivated by the
fact that weakly coupled environmental modes cannot be harnessed for
time-optimal reset~\cite{basilewitsch:2017, fischer:2019}. The ancilla can be
realized by a strongly coupled environmental mode~\cite{ShaliboPRL10,
ReichSciRep15} but also by another engineered quantum system~\cite{magnard:2018,
Ticozzi2014,Ticozzi_2017, Lau2019}. In that scenario, both qubit and ancilla are
weakly coupled to the larger thermal environment, which allows the ancilla to
equilibrate after being decoupled from the qubit.}

\change{We first consider a two-level ancilla and identify} the necessary
conditions on the qubit-ancilla time evolution operator $\op{U}$ to allow for
purification of the qubit. Employing the Cartan decomposition of
$\mathrm{SU}(4)$, every element $\op{U} \in \mathrm{SU}(4)$ can be written
as~\cite{zhang:2003}
\begin{align} \label{eq:kak-decomp}
  \op{U}
  =
  \op{K} \op{A} \op{K}',
  \qquad
  \op{A}
  =
  \exp\left\{%
    \frac{\im}{2} \sum_{k=1}^3 c_{k}
    \left(\op{\sigma}_k \otimes \op{\sigma}_k\right)
  \right\},
\end{align}
with $\op{K},\op{K}' \in \mathrm{SU}(2) \otimes \mathrm{SU}(2)$ and
$\op{\sigma}_k$ the usual Pauli matrices. This representation allows one to
separate the evolution operator $\op{U}$ into local ($\op{K}$, $\op{K}'$) and
non-local ($\op{A}$) parts. In the following, we refer to the coefficients
$c_{k} \in [0,\pi]$ as non-local (NL) coordinates and, for convenience, write
$\op{K} = \op{k}_\mathrm{S} \otimes \op{k}_\mathrm{B}$ where
$\op{k}_{\mathrm{S}}$ and $\op{k}_{\mathrm{B}}$ are local operations on qubit
and ancilla \change{(denoted by subscripts S and B for system and bath)}.
Assuming the joint initial state to be separable, $\rZ = \rS(0) \otimes \rB(0)$,
the qubit state at time $t$ is given by
\begin{equation} \label{eq:DynMap}
  \rS(t)
  =
  \tr_{\mathrm{B}} \left\{\op{U}(t) \rZ \op{U}^{\dagger}(t)\right\}
  =
  \dmap_{\rB(0)}(t) [\rS(0)] \,,
\end{equation}
where the dynamical map of the qubit, $\dmap_{\rB(0)}$, depends parametrically
on the initial ancilla state, $\rB(0)$. A necessary condition for purification
of a quantum system is non-unitality~\footnote{a map is called unital if it maps
the identity onto itself} of its dynamical map~\cite{lorenzo:2013}. In order to
check unitality in Eq.~\eqref{eq:DynMap}, we consider the initial state
\begin{align}
  \rZ
  =
  \uop_\mathrm{S}
  \otimes
  \rB(0),
  \qquad
  \rB(0)
  =
  \change{
  \begin{pmatrix}
    p_\mathrm{B}^\mathrm{e} & \gamma_\mathrm{B} \\
    \gamma_\mathrm{B}^*     & p_\mathrm{B}^\mathrm{g}
  \end{pmatrix}\,,
  }
\end{align}
\change{where $\ p_\mathrm{B}^\mathrm{g}, p_\mathrm{B}^\mathrm{e} \in [0,1]$
denote the ancilla ground and excited state populations, and $\gamma_\mathrm{B}
\in \mathbb{C}$} its coherence. Using Eq.~\eqref{eq:kak-decomp}, we find
\begin{align} \label{eq:unital_S}
  \dmap_{\rB(0)} [\uop_\mathrm{S}]
  &=
  \tr_\mathrm{B} \left\{%
    \op{U} \left(\uop_\mathrm{S} \otimes \rB(0)\right) \op{U}^\dagger
  \right\}
  \nonumber \\
  &=
  \op{k}_\mathrm{S}
  \tr_\mathrm{B} \left\{%
    \op{A} \left(\uop_\mathrm{S} \otimes \rB'\right) \op{A}^\dagger
  \right\}
  \op{k}_\mathrm{S}^\dagger,
\end{align}
where
\begin{align*}
  \rB'
  =
  \op{k}_{\mathrm{B}}' \rB(0) \op{k}_{\mathrm{B}}'^{\dagger}
  =
  \change{
  \begin{pmatrix}
    p_\mathrm{B}^{\mathrm{e}\prime} & \gamma_\mathrm{B}' \\
    \gamma_\mathrm{B}'^*            & p_\mathrm{B}^{\mathrm{g}\prime}
  \end{pmatrix}
  }
\end{align*}
is the locally transformed ancilla state. Unitality of $\dmap_{\rB(0)}$ is
determined by the partial trace in Eq.~\eqref{eq:unital_S} since
$\op{k}_{\mathrm{S}} \uop_{\mathrm{S}} \op{k}_{\mathrm{S}}^{\dagger}
= \uop_{\mathrm{S}}$ for any $\op{k}_{\mathrm{S}}$. The partial trace yields
\begin{align} \label{eq:arhoa}
  \tr_\mathrm{B} \left\{%
    \op{A} \left(\uop_\mathrm{S} \otimes \rB'\right) \op{A}^\dagger
  \right\}
  &=
  \uop_\mathrm{S}
  + 2 \rp(\gamma_\mathrm{B}') \sin(c_2) \sin(c_3) \op{\sigma}_1
  \nonumber \\
  &\phantom{=\uop_\mathrm{S}\ }
  - 2 \ip(\gamma_\mathrm{B}') \sin(c_1) \sin(c_3) \op{\sigma}_2
  \nonumber \\
  &\phantom{=\uop_\mathrm{S}\ }
  - \change{(p_\mathrm{B}^{\mathrm{g}\prime} - p_\mathrm{B}^{\mathrm{e}\prime})}
  \sin(c_1) \sin(c_2)\op{\sigma}_3.
\end{align}
Equation~\eqref{eq:arhoa} implies that any $\op{U} \in \mathrm{SU}(4)$
with only a single non-vanishing $c_{k}$ yields
a unital map for the qubit, and purification is not possible at all. Occurrence
of two non-vanishing NL coordinates is necessary but not yet sufficient for
non-unitality of $\dmap_{\rB(0)}$ due to the dependence on $\rB'$, i.e., on the
initial ancilla state $\rB(0)$ and local operation $\op{k}_{\mathrm{B}}'$.
Non-unitality of $\dmap_{\rB(0)}$, independent of the ancilla, is guaranteed by
three non-vanishing NL coordinates~\footnote{Strictly speaking, non-unitality of
$\dmap_{\rB(0)}$ also requires an ancilla initial state $\rB \neq
\uop_\mathrm{B}/2$, but this is irrelevant since a totally mixed state of the
ancilla would not allow purification in the first place.}.

\begin{figure}[tb]
  \centering
  \includegraphics{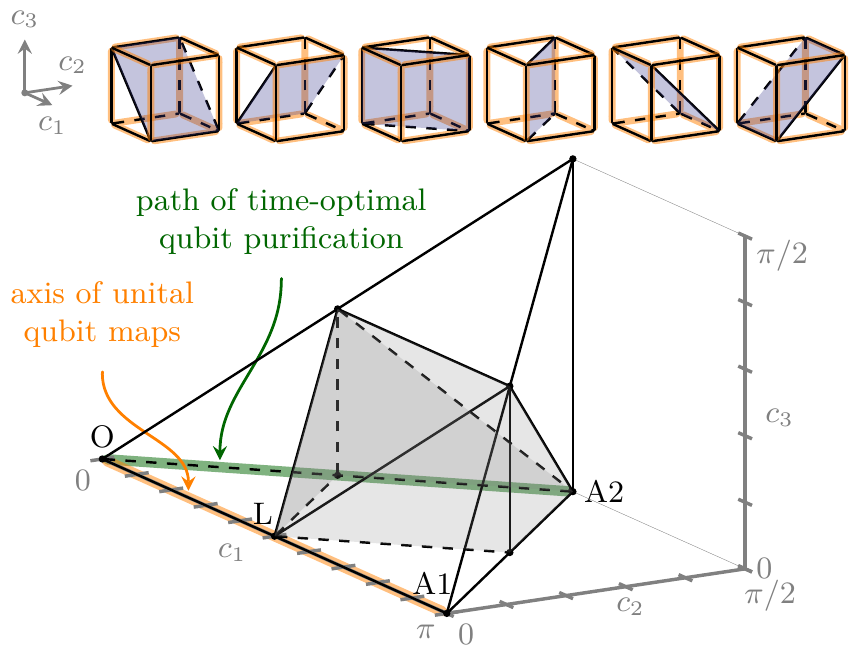}
  \caption{%
    (Color online) Symmetries (upper part) and construction of the Weyl chamber
    (lower part) for the characterization of the non-local content of any
    two-qubit operation $\op{U} \in \mathrm{SU}(4)$, cf.\
    Eq.~\eqref{eq:kak-decomp}. The shaded polyhedron within the Weyl chamber
    describes all perfectly entangling operations. The orange line highlights
    those $\op{U}$ which lead to unital maps for the qubit. The green line
    anticipates a time-optimal path for qubit purification identified below. The
    letters mark specific elements of $\mathrm{SU}(4)$ referred to in the text.
  }
  \label{fig:weyl}
\end{figure}
With this observation, we can relate non-unitality of $\dmap_{\rB(0)}$ with the
entangling capability of $\op{U}$ for the qubit-ancilla system. The latter is
best analyzed in the Weyl chamber~\cite{zhang:2003} which is a symmetry-reduced
version of the cube spanned by $c_{1}, c_{2}, c_{3} \in [0,\pi]$, obtained when
eliminating redundancies in Eq.~\eqref{eq:kak-decomp}. The six symmetries are
sketched in the upper part of Fig.~\ref{fig:weyl} with the Weyl chamber shown
below. The shaded polyhedron in its center describes all perfectly entangling
operations, and the $c_1$-axis represents all operations with at most one
non-vanishing NL coordinate. It contains one point of the polyhedron of perfect
entanglers---the point L corresponding to the gate cNOT and all gates that are
locally equivalent to it, including cPHASE. Albeit being perfect entanglers,
cNOT and cPHASE yield unital maps for the qubit. The capability of $\op{U}$ to
create entanglement between qubit and ancilla is thus a necessary but not a
sufficient condition for purification of the qubit.

Next, we determine the qubit-ancilla couplings that allow for purification of
the qubit. To this end, we write the qubit-ancilla Hamiltonian,
\begin{align} \label{eq:Ht}
  \op{H}(t)
  &=
  \op{H}_{\mathrm{S}}(t) \otimes \uop_{\mathrm{B}}
  +
  \uop_{\mathrm{S}} \otimes \op{H}_{\mathrm{B}}
  +
  \op{H}_{\mathrm{int}}
\end{align}
with $\op{H}_{\mathrm{S}}(t) = \frac{\omega_{\mathrm{S}}}{2} \op{\sigma}_{3}
+ \pulse(t) \Oc$, $\op{H}_{\mathrm{B}} = \frac{\omega_{\mathrm{B}}}{2}
\op{\sigma}_{3}$, and $\op{H}_{\mathrm{int}} = J \left(\OS \otimes \OB\right)$,
assuming control, via an external field $\pulse(t)$, only over the qubit. $J$
denotes the qubit-ancilla coupling strength~\footnote{\change{Note that it is
sufficient to consider constant couplings $J$ since time-optimal protocols will
require $J(t)$ to take its maximum value~\cite{basilewitsch:2017}}},
$\omega_{\mathrm{S}} < \omega_{\mathrm{B}}$ are the level splittings of qubit
and ancilla, and $\OS, \OB, \Oc \in \{\op{\sigma}_{1},\op{\sigma}_{2},
\op{\sigma}_{3}\}$. To relate the Hamiltonian~\eqref{eq:Ht} to the purification
condition, which is stated in terms of the number $\Nc$ of non-zero NL
coordinates of the joint qubit-ancilla time evolution, we consider the dynamical
Lie algebra $\mathcal{L}$. Its Cartan decomposition, $\mathcal{L} = \mathfrak{k}
\oplus \mathfrak{p}$, implies that $\op A$ in Eq.~\eqref{eq:kak-decomp} is
\change{an element of the group} $\exp\{\mathfrak{a}\}$ where $\mathfrak{a}$ is
\change{the} Cartan subalgebra, i.e., \change{the} maximal Abelian subalgebra of
$\mathfrak{p}$. For Hamiltonian~\eqref{eq:Ht} and the control task of purifying
the qubit, it turns out that \change{$\Nc
= \mathrm{dim}\{\mathfrak{a}\}$~\footnote{Note that the equality of
$\mathrm{dim}\{\mathfrak{a}\}$ and $\Nc$ \change{holds for Hamiltonians of the
form Eq.~\eqref{eq:Ht}, but not in general.}}. Since $\mathfrak{a}$ can be
determined entirely from the Hamiltonian, without any knowledge of the actual
dynamics, $\Nc$ is readily obtained. It is thus straightforward to identify the
combinations $\OS, \OB, \Oc \in \{\op{\sigma}_{1}, \op{\sigma}_{2},
\op{\sigma}_{3}\}$ for which $\Nc \geq 2$:} Out of the $3\times 3\times 3$
possible combinations of $\OS, \OB, \Oc$, only 16 have a Cartan subalgebra
$\mathfrak{a}$ of dimension 2, allowing for purification of the
qubit~\footnote{\change{Note that $\Nc\leq 2$ for Hamiltonian~\eqref{eq:Ht} but
this is sufficient for purification together with the assumption of a thermal
initial ancilla state, i.e., $\gamma_\mathrm{B} = 0$. Then $\gamma_B^\prime$ has
to be zero as well since the local operations $\op{k}_\mathrm{B}^\prime$ for
Hamiltonian~\eqref{eq:Ht} are generated only by $\uop_\mathrm{S} \otimes
\uop_\mathrm{B}$ and $\uop_\mathrm{S} \otimes \sigma_3$ which do not change the
ancilla coherence. $\gamma_B^\prime=0$ implies non-unitality according to
Eq.~\eqref{eq:arhoa}}}. Table~SI in the supplemental material (SM)~\cite{supp}
summarizes the resulting dynamical Lie algebras and presents possible choices
for $\mathfrak{a}$.

\change{After identifying the cases, in which Hamiltonian~\eqref{eq:Ht} allows
for purification of the qubit, we seek to derive the fields $\pulse(t)$ which
reset the qubit in minimum time. This requires consideration not only of the
generators, but also of the full dynamics} We assume a separable initial state
with the ancilla in thermal equilibrium, i.e., without ancilla coherence,
\begin{align}\label{eq:rho0}
  \rZ
  =
  \begin{pmatrix}
    p_{\mathrm{S}}^{\mathrm{e}} & \gamma_{\mathrm{S}} \\
    \gamma_{\mathrm{S}}^{*}     & p_{\mathrm{S}}^{\mathrm{g}}
  \end{pmatrix}
  \otimes
  \begin{pmatrix}
    p_{\mathrm{B}}^{\mathrm{e}} & 0 \\
    0                           & p_{\mathrm{B}}^{\mathrm{g}}
  \end{pmatrix}\,.
\end{align}
\change{Before stating the general result, we present three observations based
on numerical simulations of all 16 cases. (i) For a constant resonant field, the
minimum time, $T_{\mathrm{min}}$, to achieve maximal qubit purity is independent
of the initial qubit state, $\rS(0)$. Such a field puts qubit and ancilla into
resonance, i.e., $\pulse$ is chosen such that $\lambda_{1} - \lambda_{0}
= \omega_{\mathrm{B}}$ where $\lambda_{0}< \lambda_{1}$ are the field-dressed
eigenvalues of the qubit Hamiltonian $\op{H}_{\mathrm{S}}$. This generalizes the
results of Ref.~\cite{basilewitsch:2017} to the remaining 15 cases where
purification is possible. (ii) The specific value of $T_{\mathrm{min}}$ depends
on the type of qubit-ancilla interaction and local control, i.e., on the choice
of $\OS$, $\OB$, $\Oc$. (iii) When allowing for fully time-dependent control
fields $\pulse(t)$, the maximal qubit purity cannot be reached faster than
$T_{\mathrm{min}}$. For times $\tau < T_{\mathrm{min}}$, numerical optimizations
using Krotov's method~\cite{konnov:1999, reich:2012} lead to an upper bound for
the qubit purity $\mathcal{P}_{\mathrm{S}}(\tau)$, given in terms of the
evolution of the thermal initial state $\rS(0)$. The optimally shaped fields
saturating the bound depend both on the initial state and the choice of $\OS$,
$\OB$, $\Oc$.}

\begin{figure}[tb]
  \centering
  \includegraphics{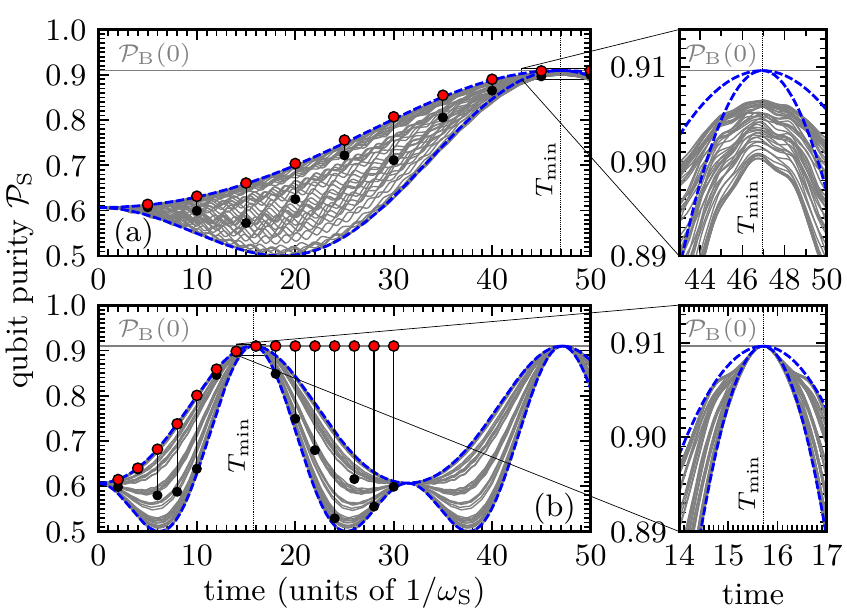}
  \caption{%
    (Color online) \change{The time evolution of the qubit purity reveals the
    minimum time for qubit reset -- shown here for the examples
    $\op{H}_\mathrm{int} = J(\op{\sigma}_1 \otimes \op{\sigma}_1)$ with
    $\op{H}_\mathrm{c}(t) = \pulse(t)\,\op{\sigma}_1$ (a) and
    $\op{H}_\mathrm{c}(t) = \pulse(t)\,\op{\sigma}_3$ (b). The gray area
    corresponds to the superimposed evolutions} of fifty different initial
    states $\rZ = \rS(0) \otimes \rB(0)$ with $\rB(0)$ the thermal equilibrium
    state and $\rS(0)$ sampled randomly under the condition of identical purity.
    The red (black) dots indicate the values for
    $\mathcal{P}_{\mathrm{S}}(\tau)$ obtained with optimized (constant resonant)
    $\pulse(t)$. \change{The dashed blue lines indicate the upper and lower
    bounds of $\mathcal{P}_{\mathrm{S}}$ predicted by Eq.~\eqref{eq:PS}.
    $\mathcal{P}_{\mathrm{B}}(0)$ denotes the initial ancilla purity}. The
    parameters are $\omega_{\mathrm{S}} = 1$, $\omega_{\mathrm{B}} = 3$,
    $J=0.1$, and inverse temperature $\beta=1$.
  }
  \label{fig:pur}
\end{figure}
\change{%
We discuss these observations in more detail for $\OS = \OB = \op{\sigma}_{1}$
with (a) $\Oc = \op{\sigma}_{1}$ and (b) $\Oc = \op{\sigma}_{3}$, \change{cf.
Fig.~\ref{fig:pur}. These two examples are paradigmatic, with all other
combinations of $\OS$, $\OB$, and $\Oc$ yielding identical results (data not
shown).}} In case (b), we know that time-optimal reset is achieved with
a constant field $\pulse(t) = \pulse$ that puts qubit and ancilla into
resonance~\cite{basilewitsch:2017}. A constant resonant field therefore seems to
be a suitable pilot for all numerical optimizations. For all initial states of
the qubit (randomly chosen under the condition of identical purity), a maximum
in the time evolution of the purity occurs at roughly the same point in time,
$T_{\mathrm{min}}$, when applying the constant resonant field, cf.
Fig.~\ref{fig:pur}. Using optimal control theory, we find field shapes
$\pulse(t)$, $t \in [0,\tau]$, that maximize $\mathcal{P}_{\mathrm{S}}(\tau)$ at
a given time $\tau$. For a few choices of $\tau$, the black and red dots in
Fig.~\ref{fig:pur} compare the purities obtained with constant resonant and
optimized fields~\footnote{Note that for the optimizations in
Fig.~\ref{fig:pur}(b), a second control, coupling to $\Oc = \op{\sigma}_{1}$,
has been utilized, since $\Oc = \op{\sigma}_{3}$ alone is not sufficient for
modifying the qubit coherence.}. In all cases with $\tau\neq T_{\mathrm{min}}$,
the optimized fields improve the purity compared to the constant resonant field.
For $\tau = T_{\mathrm{min}}$, the purities coincide (since the constant
resonant field is already an optimal solution). For $\tau < T_{\mathrm{min}}$,
there exists an upper bound for $\mathcal{P}_{\mathrm{S}}(\tau)$, which is
attained by the constant resonant field for a coherence-free initial qubit state
\change{with $p_{\mathrm{S}}^{\mathrm{g}} > p_{\mathrm{S}}^{\mathrm{e}}$}.
\change{This bound can be proven rigorously~\cite{Ticozzi2014, supp}.}
\change{While the individual optimized fields depend on $\tau$ and, for
$\tau\neq T_{\mathrm{min}}$, on the initial state of the qubit,} they all
exhibit a strong off-resonant initial peak which, if $\gamma_S\neq 0$, rotates
the coherence into population \change{or, if $p_{\mathrm{S}}^{\mathrm{g}}
< p_{\mathrm{S}}^{\mathrm{e}}$, inverts the qubit populations,} before swapping
the purities with the resonant protocol~\cite{basilewitsch:2017}. When allowing
for times $\tau > T_{\mathrm{min}}$, a purity swap between qubit and ancilla
remains the optimal strategy, with the corresponding fields being more complex
than the constant resonant solution for $\tau = T_{\mathrm{min}}$.

Based on these numerical results, we conjecture that time-optimal purification
requires one to choose a constant and resonant field, $\pulse(t)=\pulse$ such
that $\lambda_{1} - \lambda_{0} = \omega_{\mathrm{B}}$, independent of $\OS$,
$\OB$ and $\Oc$. For this choice of $\pulse(t)$, an exact closed-form expression
for the joint time evolution operator can be approximated to yield an expression
for the time evolution of the qubit purity, cf. SM~\cite{supp}. For the example
of case (a), it reads
\begin{eqnarray}
\label{eq:PS}
  \mathcal{P}_{\mathrm{S}}(t)
  &\approx&
  \Big[
      p_{\mathrm{S}}^{\mathrm{g}} p_{\mathrm{B}}^{\mathrm{g}}
    + p_{\mathrm{S}}^{\mathrm{g}} p_{\mathrm{B}}^{\mathrm{e}} \cos^{2}(\eta t)
    + p_{\mathrm{S}}^{\mathrm{e}} p_{\mathrm{B}}^{\mathrm{g}} \sin^{2}(\eta t)
  \Big]^{2}
  \nonumber\\
  &&
  +
  \Big[
      p_{\mathrm{S}}^{\mathrm{e}} p_{\mathrm{B}}^{\mathrm{e}}
    + p_{\mathrm{S}}^{\mathrm{g}} p_{\mathrm{B}}^{\mathrm{e}} \sin^{2}(\eta t)
    + p_{\mathrm{S}}^{\mathrm{e}} p_{\mathrm{B}}^{\mathrm{g}} \cos^{2}(\eta t)
  \Big]^{2}
  \nonumber \\
  &&
  +
  2 |\gamma_{\mathrm{S}}|^{2} \cos^{2}(\eta t)
\end{eqnarray}
with
\begin{align} \label{eq:eta}
  \eta^{2}
  =
  J^{2} + \frac{\omega_{\mathrm{B}}}{2}
  \left(\omega_{\mathrm{B}} - \sqrt{\omega_{\mathrm{B}}^{2} + 4 B^{2}}\right),
  \;
  B^{2}
  =
  J^{2} \frac{%
    \omega_{\mathrm{B}}^{2} - \omega_{\mathrm{S}}^{2}
  }{%
    \omega_{\mathrm{B}}^{2}
  }.
\end{align}
Maximizing $\mathcal{P}_{\mathrm{S}}(t)$ yields \change{an approximated} minimum
time for purification, $T_{\mathrm{min}} = \pi/(2\eta)$, independent of the
initial state. This explains why $\mathcal{P}_{\mathrm{S}}(t)$ displays
a perfect swap of qubit and ancilla purity at $T_{\mathrm{min}}$ for all
evolutions in Fig.~\ref{fig:pur}(a). For all other Hamiltonians which allow for
purification, we obtain the same result for $\mathcal{P}_{\mathrm{S}}(t)$,
Eq.~\eqref{eq:PS}, but with different expressions for $\eta$ and
$B$~\cite{supp}. \change{For example, in case (b) $B=0$ and hence $\eta=J$ leads
to $T_{\mathrm{min}} = \pi/(2J)$, which is much shorter than $T_{\mathrm{min}}$
in case (a) where $B \neq 0$, cf. Fig.~\ref{fig:pur}.}

Overall, we find three distinct minimal reset times which can be linked to
a single quantity $A$ with $T_{\mathrm{min}} \approx \pi/(2|A|)$. \change{The
parameter} $A$ can be obtained analytically from the Hamiltonian and is
determined by the parameters $J$, $\omega_{\mathrm{S}}$ and
$\omega_{\mathrm{B}}$ for all possible combinations of $\OS, \OB, \Oc \in
\{\op{\sigma}_{1}, \op{\sigma}_{2}, \op{\sigma}_{3}\}$, cf. SM~\cite{supp}. The
minimal times depending on the choice of $\OS, \OB, \Oc$ are~\cite{supp}:
$T^{(1)}_{\mathrm{min}} = \pi / (2J)$, $T^{(2)}_{\mathrm{min}} = \pi / (2J)
\cdot \omega_{\mathrm{B}} / \omega_{\mathrm{S}}$, and $T^{(3)}_{\mathrm{min}}
= \pi / (2J) \cdot \omega_{\mathrm{B}} / \sqrt{\omega_{\mathrm{B}}^{2}
- \omega_{\mathrm{S}}^{2}}$, see also Table~\ref{tab:Tmin_exp}. Among these,
$T^{(1)}_{\mathrm{min}}$ represents the global minimum, \change{as we prove
rigorously in the SM~\cite{supp}, using the Time-optimal tori theorem for
$\mathrm{SU}(4)$~\cite{khaneja:2001}}. \change{ Remarkably, we find that, for
qubit reset, the global minimum $T^{(1)}_{\mathrm{min}}$ can be attained with
a single control field acting on the qubit only, whereas the Time-optimal tori
theorem assumes full controllability over system and ancilla.}
\change{When allowing $\OS$, $\OB$ and $\Oc$ to be arbitrary elements of
$\mathfrak{su}(2)=\mathrm{span}\{\op{\sigma}_{1}, \op{\sigma}_{2},
\op{\sigma}_{3}\}$ including superpositions of Pauli operators, we find the
respective minimum times for the purity swap to be lower bounded by
$T^{(1)}_{\mathrm{min}}$, cf.  SM~\cite{supp}.}

\begin{table}
  \centering
  \caption{%
    Minimum reset times $T_{\mathrm{min}}$ for two coupled superconducting
    qubits~\cite{PRL.94.240502} with three sets of experimental parameters
    (set 1:
    $\omega_{\mathrm{S}}/(2\pi) = 12.8$~GHz,
    $\omega_{\mathrm{B}}/(2\pi) = 16.1$~GHz,
    $J/(2\pi) = 65$~MHz;
    set 2:
    $\omega_{\mathrm{S}}/(2\pi) = 9.8$~GHz,
    $\omega_{\mathrm{B}}/(2\pi) = 16.1$~GHz,
    $J/(2\pi) = 200$~MHz;
    set 3:
    $\omega_{\mathrm{S}}/(2\pi) = 15.8$~GHz,
    $\omega_{\mathrm{B}}/(2\pi) = 16.1$~GHz,
    $J/(2\pi) = 25$~MHz~\cite{PRB.81.134507}).
  }
  \begin{tabular*}{\linewidth}{c @{\extracolsep{\fill}} ccccc}
    \hline
    $\OS \otimes \OB$ & $\Oc$
    & \makecell{$T_{\mathrm{min}}$ \\ (set 1)}
    & \makecell{$T_{\mathrm{min}}$ \\ (set 2)}
    & \makecell{$T_{\mathrm{min}}$ \\ (set 3)} \\
    \hline
    $\xx$ & $\op{\sigma}_{1}$ & $191.0$~ns & $81.1$~ns & $402.3$~ns \\
    $\xx$ & $\op{\sigma}_{2}$ & $151.8$~ns & $49.3$~ns & $394.8$~ns \\
    $\xx$ & $\op{\sigma}_{3}$ & $151.8$~ns & $49.3$~ns & $394.8$~ns \\
    \hline
    $\xy$ & $\op{\sigma}_{1}$ & $191.0$~ns & $81.1$~ns & $402.3$~ns \\
    $\xy$ & $\op{\sigma}_{2}$ & $151.8$~ns & $49.3$~ns & $394.8$~ns \\
    $\xy$ & $\op{\sigma}_{3}$ & $151.8$~ns & $49.3$~ns & $394.8$~ns \\
    \hline
    $\yx$ & $\op{\sigma}_{1}$ & $151.8$~ns & $49.3$~ns & $394.8$~ns \\
    $\yx$ & $\op{\sigma}_{2}$ & $191.0$~ns & $81.1$~ns & $402.3$~ns \\
    $\yx$ & $\op{\sigma}_{3}$ & $151.8$~ns & $49.3$~ns & $394.8$~ns \\
    \hline
    $\yy$ & $\op{\sigma}_{1}$ & $151.8$~ns & $49.3$~ns & $394.8$~ns \\
    $\yy$ & $\op{\sigma}_{2}$ & $191.0$~ns & $81.1$~ns & $402.3$~ns \\
    $\yy$ & $\op{\sigma}_{3}$ & $151.8$~ns & $49.3$~ns & $394.8$~ns \\
    \hline
    $\zx$ & $\op{\sigma}_{1}$ & $250.3$~ns & $62.2$~ns & $2054.6$~ns \\
    $\zx$ & $\op{\sigma}_{2}$ & $250.3$~ns & $62.2$~ns & $2054.6$~ns \\
    \hline
    $\zy$ & $\op{\sigma}_{1}$ & $250.3$~ns & $62.2$~ns & $2054.6$~ns \\
    $\zy$ & $\op{\sigma}_{2}$ & $250.3$~ns & $62.2$~ns & $2054.6$~ns \\
    \hline
  \end{tabular*}
  \label{tab:Tmin_exp}
\end{table}
Table~\ref{tab:Tmin_exp} shows exemplary minimum reset times $T_{\mathrm{min}}$
for a physical realization of two coupled superconducting
qubits~\cite{PRL.94.240502,PRB.81.134507}, illustrating the role of device
design in terms of the choice of $\OS$, $\OB$, and $\Oc$ as well as the
parameters \change{$\omega_\mathrm{S},\ \omega_\mathrm{B}$ and $J$}. In case of
the optimal choice of $\OS$, $\OB$, and $\Oc$ such that
$T_{\mathrm{min}}=T^{(1)}_{\mathrm{min}}$ (independently of $\omega_{B}$),
$\omega_{B}$ should be chosen as large as possible to increase the maximum
achievable purity. \change{An option to effectively enlarge $\omega_{B}$ is to
utilize ancilla levels above the two-level subspace. This raises the question
whether our findings on minimum protocol duration and maximum reset fidelity
hold also for ancillas with Hilbert space dimension larger than two.}

\change{Numerical simulations for a three-level and four-level ancilla suggest
that the reset time for ancillas with Hilbert space dimension larger than two is
still lower bounded by $T^{(1)}_{\mathrm{min}} = \pi/(2J)$~\cite{supp}. In
contrast, the maximally reachable qubit purity increases and can be brought
close to one for ancillas with Hilbert space dimension $d_\mathrm{B}\ge 4$,
provided half of the eigenvalues of the initial ancilla state are
small~\cite{supp}. If we consider, for example, an ancilla with equidistant
energy levels ($\omega_\mathrm{B}=3$) and thermal initial states on qubit
($\omega_\mathrm{S} = 1$) and ancilla with $\beta = 1$, the maximal achievable
purity increases from $\mathcal{P}\approx 0.970$ for $d_\mathrm{B} = 3$ to
$\mathcal{P} \approx 0.995$ for $d_\mathrm{B} = 4$, while $\mathcal{P} \approx
0.905$ for a two-level ancilla. These conditions can easily be realized by
utilizing higher excited levels of transmon qubits or superconducting
resonators. Such ancillas would also provide for complete
controllability~\cite{geerlings:2013, magnard:2018, Egger2018} which may be
required to realize the spectral reshuffling of the joint initial state that
implements an optimal reset dynamics, as we discuss in more detail in the
SM~\cite{supp}.}

To conclude, we have shown that there exists a globally minimal time (among all
Hamiltonians) to reset a qubit with maximal fidelity when making use of an
ancilla \change{and time-dependent external control fields over the qubit}.
\change{For two-level ancillas}, a time-optimal protocol ensures resonance
between qubit and ancilla and swaps their purities. The reset fidelity is then
determined by the initial ancilla purity, making it crucial to engineer
a sufficiently high ancilla purity, or, respectively, low ancilla temperature.
\change{Due to its non-trivial dependence on the effective qubit-ancilla
coupling strength,} there exists an optimal choice for qubit-ancilla interaction
and type of local control for the time-optimal reset. Thanks to the Cartan
decomposition of SU(4), this choice can be determined at the level of the
algebra, i.e., the Hamiltonian, and does not require knowledge of the actual
reset dynamics. \change{For ancillas with Hilbert space dimension larger than
three, the qubit purity can be brought arbitrarily close to one by proper
spectral reshuffling of the joint qubit ancilla state, provided at least half of
the ancilla levels have negligible population. The corresponding experimental
conditions on ancilla state and control are easily met by e.g. current
superconducting qubit technology}. Our results provide the guiding principles
for device design in order to realize the fastest and most accurate protocol for
qubit reset in a given QI architecture.

\begin{acknowledgments}
  \change{We would like to thank Hoi Kwan Lau, Aashish Clerk, and Francesco
  Ticozzi for helpful discussions.} Financial support from the
  Volkswagenstiftung Project No. 91004 and the ANR-DFG research program COQS
  (ANR-15-CE30-0023-01, DFG COQS Ko 2301/11-1) is gratefully acknowledged.
\end{acknowledgments}


%

\end{document}


\title{%
  Supplemental Material: Fundamental Bounds on Qubit Reset
}
\author{Daniel Basilewitsch}
\affiliation{%
  Theoretische Physik, Universit\"{a}t Kassel, D-34132 Kassel, Germany
}
\affiliation{%
  Dahlem Center for Complex Quantum Systems and Fachbereich Physik, Freie
  Universit\"{a}t Berlin, Arnimallee 14, D-14195 Berlin, Germany
}

\author{Jonas Fischer}
\affiliation{%
  Theoretische Physik, Universit\"{a}t Kassel, D-34132 Kassel, Germany
}
\affiliation{%
  Dahlem Center for Complex Quantum Systems and Fachbereich Physik, Freie
  Universit\"{a}t Berlin, Arnimallee 14, D-14195 Berlin, Germany
}
\affiliation{%
  Laboratoire Interdisciplinaire Carnot de Bourgogne (ICB), Universit\'e de
  Bourgogne-Franche Comt\'e, F-21078 Dijon Cedex, France
}

\author{Daniel M. Reich}
\affiliation{%
  Theoretische Physik, Universit\"{a}t Kassel, D-34132 Kassel, Germany
}
\affiliation{%
  Dahlem Center for Complex Quantum Systems and Fachbereich Physik, Freie
  Universit\"{a}t Berlin, Arnimallee 14, D-14195 Berlin, Germany
}

\author{Dominique Sugny}
\affiliation{%
  Laboratoire Interdisciplinaire Carnot de Bourgogne (ICB), Universit\'e de
  Bourgogne-Franche Comt\'e, F-21078 Dijon Cedex, France
}

\author{Christiane P. Koch}
\affiliation{%
  Theoretische Physik, Universit\"{a}t Kassel, D-34132 Kassel, Germany
}
\affiliation{%
  Dahlem Center for Complex Quantum Systems and Fachbereich Physik, Freie
  Universit\"{a}t Berlin, Arnimallee 14, D-14195 Berlin, Germany
}

\date{\today}

\maketitle

We provide here the details of our calculations for two-level ancillas in
Sec.~\ref{s:app:twolevel} and for ancillas with larger Hilbert space dimension
in Sec.~\ref{s:app:qudits}. In particular, in Table~\ref{s:tab:dynLie}, we
present a detailed overview over the dynamical Lie algebras for the 27 possible
choices of qubit-ancilla interaction and local control in
Hamiltonian~(\axref{6}). In Sec.~\ref{s:app:qsl}, we use the Time-Optimal Tori
Theorem of Ref.~\cite{khaneja:2001} to deduce the quantum speed limit (QSL) time
for qubit reset. We present details of the derivation of Eq.~(\axref{8}) of
the main text in Sec.~\ref{s:app:ana} and discuss an extension of our model to
superpositions of Pauli operators in the interaction and control Hamiltonians in
Sec.~\ref{s:app:superpos}. For ancillas with Hilbert space dimension larger than
two, we discuss the achievable minimum time and maximum reachable purity in
Sec.~\ref{subsec:tmind3} and~\ref{subsec:purityd}, respectively.

\begin{table*}
  \centering
  \caption{%
    Dynamical Lie algebras $\mathcal{L}$, their respective Cartan
    decompositions $\mathcal{L} = \mathfrak{k} \oplus \mathfrak{p}$ and
    a possible choice for the Cartan subalgebra $\mathfrak{a} \subset
    \mathfrak{p}$ for all possible variants of Hamiltonian~(\axref{6}). The
    interaction part and the qubit control are given by $\op{H}_{\mathrm{int}}
    = J(\OS \otimes \OB)$ and $\op{H}_{\mathrm{c}}(t) = \pulse(t) \Oc$,
    respectively.
  }
  \begin{tabular*}{\linewidth}{c @{\extracolsep{\fill}} cccc}
    \hline
    $\OS \otimes \OB$ & $\Oc$ & $\mathfrak{k}$ & $\mathfrak{p}$
    & $\mathfrak{a}$ \\
    \hline
    $\xx$ & $\op{\sigma}_{1}$ & $\uz, \xu, \yu, \zu$ & $\xx, \xy, \yx, \yy, \zx, \zy$ & $\xx, \yy$ \\
    $\xx$ & $\op{\sigma}_{2}$ & $\uz, \xu, \yu, \zu$ & $\xx, \xy, \yx, \yy, \zx, \zy$ & $\xx, \yy$ \\
    $\xx$ & $\op{\sigma}_{3}$ & $\uz, \zu$           & $\xx, \xy, \yx, \yy$           & $\xx, \yy$ \\
    \hline
    $\xy$ & $\op{\sigma}_{1}$ & $\uz, \xu, \yu, \zu$ & $\xx, \xy, \yx, \yy, \zx, \zy$ & $\xx, \yy$ \\
    $\xy$ & $\op{\sigma}_{2}$ & $\uz, \xu, \yu, \zu$ & $\xx, \xy, \yx, \yy, \zx, \zy$ & $\xx, \yy$ \\
    $\xy$ & $\op{\sigma}_{3}$ & $\uz, \zu$           & $\xx, \xy, \yx, \yy$           & $\xx, \yy$ \\
    \hline
    $\xz$ & $\op{\sigma}_{1}$ & $\uz, \xu, \yu, \zu$ & $\xz, \yz, \zz$                & $\zz$      \\
    $\xz$ & $\op{\sigma}_{2}$ & $\uz, \xu, \yu, \zu$ & $\xz, \yz, \zz$                & $\zz$      \\
    $\xz$ & $\op{\sigma}_{3}$ & $\uz, \zu$           & $\xz, \yz$                     & $\xz$      \\
    \hline
    $\yx$ & $\op{\sigma}_{1}$ & $\uz, \xu, \yu, \zu$ & $\xx, \xy, \yx, \yy, \zx, \zy$ & $\xx, \yy$ \\
    $\yx$ & $\op{\sigma}_{2}$ & $\uz, \xu, \yu, \zu$ & $\xx, \xy, \yx, \yy, \zx, \zy$ & $\xx, \yy$ \\
    $\yx$ & $\op{\sigma}_{3}$ & $\uz, \zu$           & $\xx, \xy, \yx, \yy$           & $\xx, \yy$ \\
    \hline
    $\yy$ & $\op{\sigma}_{1}$ & $\uz, \xu, \yu, \zu$ & $\xx, \xy, \yx, \yy, \zx, \zy$ & $\xx, \yy$ \\
    $\yy$ & $\op{\sigma}_{2}$ & $\uz, \xu, \yu, \zu$ & $\xx, \xy, \yx, \yy, \zx, \zy$ & $\xx, \yy$ \\
    $\yy$ & $\op{\sigma}_{3}$ & $\uz, \zu$           & $\xx, \xy, \yx, \yy$           & $\xx, \yy$ \\
    \hline
    $\yz$ & $\op{\sigma}_{1}$ & $\uz, \xu, \yu, \zu$ & $\xz, \yz, \zz$                & $\zz$      \\
    $\yz$ & $\op{\sigma}_{2}$ & $\uz, \xu, \yu, \zu$ & $\xz, \yz, \zz$                & $\zz$      \\
    $\yz$ & $\op{\sigma}_{3}$ & $\uz, \zu$           & $\xz, \yz$                     & $\xz$      \\
    \hline
    $\zx$ & $\op{\sigma}_{1}$ & $\uz, \xu, \yu, \zu$ & $\xx, \xy, \yx, \yy, \zx, \zy$ & $\xx, \yy$ \\
    $\zx$ & $\op{\sigma}_{2}$ & $\uz, \xu, \yu, \zu$ & $\xx, \xy, \yx, \yy, \zx, \zy$ & $\xx, \yy$ \\
    $\zx$ & $\op{\sigma}_{3}$ & $\uz, \zu$           & $\zx, \zy$                     & $\zx$      \\
    \hline
    $\zy$ & $\op{\sigma}_{1}$ & $\uz, \xu, \yu, \zu$ & $\xx, \xy, \yx, \yy, \zx, \zy$ & $\xx, \yy$ \\
    $\zy$ & $\op{\sigma}_{2}$ & $\uz, \xu, \yu, \zu$ & $\xx, \xy, \yx, \yy, \zx, \zy$ & $\xx, \yy$ \\
    $\zy$ & $\op{\sigma}_{3}$ & $\uz, \zu$           & $\zx, \zy$                     & $\zx$      \\
    \hline
    $\zz$ & $\op{\sigma}_{1}$ & $\uz, \xu, \yu, \zu$ & $\xz, \yz, \zz$                & $\zz$      \\
    $\zz$ & $\op{\sigma}_{2}$ & $\uz, \xu, \yu, \zu$ & $\xz, \yz, \zz$                & $\zz$      \\
    $\zz$ & $\op{\sigma}_{3}$ & $\uz, \zu$           & $\zz$                          & $\zz$      \\
    \hline
  \end{tabular*}
  \label{s:tab:dynLie}
\end{table*}

\section{Two-level ancillas}
\label{s:app:twolevel}

\subsection{Quantum speed limit}
\label{s:app:qsl}
We construct the globally minimal time to reset a qubit using a two-level
ancilla, starting from a fully controllable two-qubit system and employing the
Time-optimal tori theorem~\cite{khaneja:2001}. Assuming local control over both
qubit and ancilla allows us to derive the minimal time from the Time-Optimal
Tori Theorem. Since our model in the main text does not include control over the
ancilla, the mininum time derived here from the Time-optimal tori theorem will
be a lower bound for qubit reset. We show that, when relaxing the
controllability assumption to local control over the system qubit only, the
bound is tight and can be attained for specific choices of the qubit-ancilla
Hamiltonian.

For the sake of completeness, we first recall Theorems 2 and 10 of
Ref.~\cite{khaneja:2001} in the notation of our study. We consider a system
qubit coupled to a bath qubit (ancilla), described by the Hamiltonian
\begin{equation}\label{eq1}
  \op H=\op H_\mathrm{S}\otimes \openone_\mathrm{B}+\openone_\mathrm{S}\otimes
  \op H_\mathrm{B}+\op H_{\mathrm{int}}\,,
\end{equation}
with
\begin{eqnarray*}
  \op H_\mathrm{S}&=&u_{1,\mathrm{S}}\,\op\sigma_1+u_{2,\mathrm{S}}\,\op\sigma_2
  + u_{3,\mathrm{S}}\,\op{\sigma}_3\\
  \op H_\mathrm{B}&=&u_{1,\mathrm{B}}\,\op\sigma_1+u_{2,\mathrm{B}}\,\op\sigma_2
  + u_{3,\mathrm{B}}\,\op{\sigma}_3 \\
  \op H_{\mathrm{int}}&=&J\,\op\sigma_k\otimes \op\sigma_l,
\end{eqnarray*}
where $\op\sigma_k,~\op\sigma_l\in \{\op\sigma_1,\op\sigma_2,\op\sigma_3\}$, and
$J$ is a constant coupling strength. In contrast to our study,
Ref.~\cite{khaneja:2001} assumes complete controllability, i.e., local controls
on qubit and ancilla generating the subgroup $K=\mathrm{SU}(2) \otimes
\mathrm{SU}(2)$. The dynamical Lie algebra corresponding to Eq.~\eqref{eq1} is
thus $\mathfrak{su}(4)$. No limitations on the maximum intensity of the control
fields $u_{i,\mathrm{S}}$ and $u_{i,\mathrm{B}}$ are imposed so that any local
operation can be realized in arbitrarily short time which can be neglected
compared to the timescale of the interaction $1/J$.
\begin{theorem}[Time-Optimal Tori Theorem~\cite{khaneja:2001}]\label{theo}
  For the system described by Eq.~\eqref{eq1}, the minimum time to generate
  a unitary propagator $\op U_\mathrm{F}\in \mathrm{SU}(4)$ corresponds to the
  smallest value of $\sum_k c_k$, where $c_k\in [0,\pi]$, such that
  $$
    \op U_\mathrm{F}
    =
    \op{K} e^{\frac{\im}{2}\sum_k c_k\op\sigma_k\otimes\op\sigma_k}\op{K}',
  $$
  with $\op{K}, \op{K}' \in K$. The minimum time $T$ is given by
  $T=\frac{1}{2J}\sum_k c_k$.
\end{theorem}
The proof of Theorem~\ref{theo}~\cite{khaneja:2001} is based on the same Cartan
decomposition of $\mathrm{SU}(4)$ as used in the main text. A minimum-time
control strategy can be derived from Theorem~\ref{theo} by expressing the
propagator $\op U_\mathrm{F}$ as
\begin{align*}
  \op U_\mathrm{F} &=
  \op K
  \left(%
    \prod_{k = 1}^{3} e^{\frac{\im}{2} c_k \op \sigma_k \otimes \op \sigma_k}
  \right)
  \op K'\\
  &=
  \op K'_1 e^{\frac{\im}{2J}c_1 \op H_{\mathrm{int}}}
  \op K'_2 e^{\frac{\im}{2J}c_2 \op H_{\mathrm{int}}}
  \op K'_3 e^{\frac{\im}{2J}c_3 \op H_{\mathrm{int}}} \op K'_4\,,
\end{align*}
where $\op K'_1,\op K'_2,\op K'_3,\op K'_4\in K$ are chosen such that they
rotate $\op H_\mathrm{int}$ into the corresponding term of $\op \sigma_k
\otimes \op \sigma_k$.
The control protocol thus consists in a series of local pulses (which take no
time) and field-free evolutions under the interaction Hamiltonian $\op
H_{\mathrm{int}}$ (of duration $c_k/(2J)$). We now apply Theorem~\ref{theo} to
qubit reset, taking the initial state as $\rho(0)=\openone/2 \otimes
\rho_\mathrm{B}(0)$ with $\mathcal{P}_{\mathrm{B}}(0) = \tr
\left\{\rho_\mathrm{B}^2(0)\right\}$ the initial ancilla purity. To simplify
the description, we assume the qubit to be initially in the maximally mixed
state which is the state requiring the largest entropy/purity change. The
initial ancilla state is written as
$$
  \rho_\mathrm{B}(0)
  =
  \begin{pmatrix}
    p_{\mathrm{B}}^{\mathrm{e}} & \gamma_{\mathrm{B}} \cr
    \gamma_{\mathrm{B}}^* & p_{\mathrm{B}}^{\mathrm{g}}
  \end{pmatrix}
$$
with $\mathcal{P}_{\mathrm{B}}(0) = (p_{\mathrm{B}}^{\mathrm{g}})^{2}
+ (p_{\mathrm{B}}^{\mathrm{e}})^{2} + 2|\gamma_{\mathrm{B}}|^2$. Denoting the
evolution operator that realizes the qubit reset in minimum time,
$T_{\mathrm{min}}$, by $\op U_\mathrm{F}$ and using Theorem~\ref{theo}, we
obtain
$$
  \rho_\mathrm{S}(T_{\mathrm{min}})
  =
  \tr_\mathrm{B}\left\{\op U_\mathrm{F}\rho(0)\op U_\mathrm{F}^\dagger\right\}.
$$
Inserting $\op U_\mathrm{F}$ from Theorem~\ref{theo} leads to
\begin{eqnarray*}
  \rho_\mathrm{S}(T_{\mathrm{min}})
  &=&
  \openone/2+\rp(\gamma_{\mathrm{B}}')\sin(c_2)\sin(c_3)\,\op\sigma_1\\
  &&-\ip(\gamma_{\mathrm{B}}')\sin(c_1)\sin(c_3)\,\op\sigma_2\\
  & & - \frac{1}{2} \left(p_{\mathrm{B}}^{\mathrm{g} \prime} -
  p_{\mathrm{B}}^{\mathrm{e} \prime}\right)\sin(c_1)\sin(c_2)\,\op\sigma_3\,,
\end{eqnarray*}
where $p_{\mathrm{B}}^{\mathrm{g} \prime}, p_{\mathrm{B}}^{\mathrm{e} \prime}$
and $\gamma_{\mathrm{B}}'$ are the matrix elements of $\rho_\mathrm{B}(0)$ in
a new basis,
$$
  \begin{pmatrix}
    p_{\mathrm{B}}^{\mathrm{e} \prime} & \gamma_{\mathrm{B}}' \cr
    \gamma_{\mathrm{B}}'^* & p_{\mathrm{B}}^{\mathrm{g} \prime}
  \end{pmatrix}
  =\op{k}_\mathrm{B}
  \begin{pmatrix}
    p_{\mathrm{B}}^{\mathrm{e}} & \gamma_{\mathrm{B}} \cr
    \gamma_{\mathrm{B}}'^* & p_{\mathrm{B}}^{\mathrm{g}}
  \end{pmatrix}
  \op{k}_\mathrm{B}^\dagger
$$
with $\op{k}_\mathrm{B}$ being a unitary local operation on the ancilla. The
coefficients $p_{\mathrm{B}}^{\mathrm{g} \prime}, p_{\mathrm{B}}^{\mathrm{e}
\prime}$ and $\gamma_{\mathrm{B}}'$ fulfill the constraint
\begin{equation}\label{eqbeta}
  \mathcal{P}_{\mathrm{B}}(0) = (p_{\mathrm{B}}^{\mathrm{g} \prime})^{2} +
  (p_{\mathrm{B}}^{\mathrm{e} \prime})^{2} + 2|\gamma_{\mathrm{B}}'|^2,
\end{equation}
since local operations cannot change the ancilla purity. The qubit purity at
time $T_\mathrm{min}$ becomes
\begin{align}
  \mathcal{P}_{\mathrm{S}}(T_{\mathrm{min}})
  &=
  \tr\left\{\rho_\mathrm{S}^2(T_{\mathrm{min}})\right\}
  \notag \\
  &=
  \frac{1}{2}+\left((p_{\mathrm{B}}^{\mathrm{g} \prime})^{2}
  + (p_{\mathrm{B}}^{\mathrm{e} \prime})^{2} - \frac{1}{2}\right)
  \sin^2(c_1)\sin^2(c_2)
  \nonumber \\
  &\quad
  +2\,\rp(\gamma_{\mathrm{B}}')^2\sin^2(c_2)\sin^2(c_3)
  \nonumber \\
  &\quad
  +2\,\ip(\gamma_{\mathrm{B}}')^2\sin^2(c_1)\sin^2(c_3)
  \label{maxpur}
\end{align}
and is upper bounded by
\begin{align}
  \notag
  \mathcal{P}_{\mathrm{S}}(T_{\mathrm{min}})
  &\leq
  (p_{\mathrm{B}}^{\mathrm{g} \prime})^{2}
  +
  (p_{\mathrm{B}}^{\mathrm{e} \prime})^{2}
  +
  2\,\rp (\gamma_{\mathrm{B}}')^2+2\,\ip(\gamma_{\mathrm{B}}')^2
  \\
  \notag
  &=
  \mathcal{P}_{\mathrm{B}}(0),
\end{align}
i.e., the maximum qubit purity $\mathcal{P}_{\mathrm{S}}(T_{\mathrm{min}})$ is
the initial purity $\mathcal{P}_{\mathrm{B}}(0)$ of the
ancilla~\cite{Ticozzi2014}.

This upper bound can be attained, for instance, if $c_1=c_2=c_3=\frac{\pi}{2}$.
Employing Theorem~\ref{theo}, this leads to a control time $T=\frac{3\pi}{4J}$.
Note that this control strategy works for any value of
$p_{\mathrm{B}}^{\mathrm{g} \prime}, p_{\mathrm{B}}^{\mathrm{e} \prime}$ and
$\gamma_{\mathrm{B}}^\prime$. However, this time is not the minimum time and
shorter durations can be found under certain assumptions on
$p_{\mathrm{B}}^{\mathrm{g} \prime}, p_{\mathrm{B}}^{\mathrm{e} \prime}$, and
$\gamma_{\mathrm{B}}^\prime$, or rather on the local transformation
$\op{k}_\mathrm{B}$, as we show next.

Direct analysis of Eq.~\eqref{maxpur} reveals that the maximum purity can be
attained in time $T=\pi/(2J)$ for three symmetric processes in which one of the
three parameters $c_k$ is zero and the other two equal $\pi/2$. Whether it is
possible to utilize one of these strategies depends on the local transformation
$\op k_\mathrm{B}$ of the ancilla. If, for example,
$\gamma_{\mathrm{B}}^\prime=0$ and hence $\mathcal{P}_\mathrm{B}(0)
= (p_{\mathrm{B}}^{\mathrm{g} \prime})^2 +(p_{\mathrm{B}}^{\mathrm{e}
\prime})^2$, setting $c_3 = 0$ yields a purification time $T=\pi/(2J)$. These
requirements are fulfilled if the ancilla is initially in thermal equilibrium
and the local transformation $\op k_\mathrm{B}$ is generated only by $\op
\sigma_3$. Similarly, we find for $\rp(\gamma_{\mathrm{B}}^\prime)^2
= \frac{\mathcal{P}_{\mathrm{B}}(0)}{2}-\frac{1}{4}$ that $c_1=0$ allows for
fast purification, and with $c_2 = 0$ this time can be achieved if
$\ip(\gamma_{\mathrm{B}}^\prime)^2
= \frac{\mathcal{P}_{\mathrm{B}}(0)}{2}-\frac{1}{4}$. We prove in
Theorem~\ref{theo2} that these protocols are the time-optimal strategies for
qubit reset with the Hamiltonian~\eqref{eq1}.
\begin{theorem}\label{theo2}
  Given Eq.~\eqref{eq1}, the minimum time to reset the qubit and reach the
  maximum possible purity from a maximally mixed state is
  $T_{\mathrm{min}}=\pi/(2J)$. The time-optimal control strategies correspond to
  the cases where two of the $c_k$ in $\op U_\mathrm{F}$ are equal to $\pi/2$
  and the third one is zero. The control protocol depends on the local operation
  on the ancilla characterized by the parameters $p_{\mathrm{B}}^{\mathrm{g}
  \prime}, p_{\mathrm{B}}^{\mathrm{e} \prime}$ and $\gamma_{\mathrm{B}}^\prime$
  and is optimized to attain maximum purity.
\end{theorem}
\begin{proof}
Using Eq.~\eqref{eqbeta}, we find for the qubit purity
\begin{align}\label{eqpurity}
  \mathcal{P}_{\mathrm{S}}(T_{\mathrm{min}})
  &=
  \frac{1}{2}
  +
  \left(\mathcal{P}_{\mathrm{B}}(0)-\frac{1}{2}\right)\sin^2(c_1)\sin^2(c_2) \nonumber \\
  &\quad +2\rp(\gamma_{\mathrm{B}}')^2\sin^2(c_2)(\sin^2(c_3)-\sin^2(c_1)) \nonumber \\
  &\quad +2\ip(\gamma_{\mathrm{B}}')^2\sin^2(c_1)(\sin^2(c_3)-\sin^2(c_2))\,,
\end{align}
i.e., the qubit purity can be interpreted as a function
$F_{\gamma_{\mathrm{B}}'}$ of $(c_1,c_2,c_3)\in [0,\pi]^3$ parameterized by
$\gamma_{\mathrm{B}}'$. A maximum of $F_{\gamma_{\mathrm{B}}'}$ fulfills the
necessary conditions $\partial_{c_1}F_{\gamma_{\mathrm{B}}'}
= \partial_{c_2}F_{\gamma_{\mathrm{B}}'}
= \partial_{c_3}F_{\gamma_{\mathrm{B}}'}=0$, where $\partial_{c_k}$ denotes
the partial derivative with respect to $c_k$. We obtain the following system
of equations
\begin{widetext}
  \begin{subequations}\label{eqs}
    \begin{eqnarray}
      \sin(2c_1)\left[\left(\frac{\mathcal{P}_{\mathrm{B}}(0)}{2}-\frac{1}{4}-\rp(\gamma_{\mathrm{B}}')^2-\ip(\gamma_{\mathrm{B}}')^2\right)\sin^2(c_2)+\ip(\gamma_{\mathrm{B}}')^2\sin^2(c_3)\right]&=&0 \\
      \sin(2c_2)\left[\left(\frac{\mathcal{P}_{\mathrm{B}}(0)}{2}-\frac{1}{4}-\rp(\gamma_{\mathrm{B}}')^2-\ip(\gamma_{\mathrm{B}}')^2\right)\sin^2(c_1)+\rp(\gamma_{\mathrm{B}}')^2\sin^2(c_3)\right]&=&0 \\
      \sin(2c_3)\bigg[\rp(\gamma_{\mathrm{B}}')^2\sin^2(c_2)+\ip(\gamma_{\mathrm{B}}')^2\sin^2(c_1)\bigg]&=&0. \label{eq:third}
    \end{eqnarray}
  \end{subequations}
\end{widetext}
We study below the different extrema of $F_{\gamma_{\mathrm{B}}'}$ that
allow to achieve maximum purity in a time shorter or equal to $\pi/(2J)$,
i.e., for $c_1+c_2+c_3\leq \pi$. We do not consider the other cases.

First, we examine Eq.~\eqref{eq:third} and realize that it is satisfied if
$\sin(2 c_3)=0$, leading to $c_3 = 0$ or $c_3 = \pi/2$ (the case $c_3 = \pi$ is
not relevant). In the former case we can deduce from Eqs.~\eqref{eqpurity} and
\eqref{eqs} that $c_1=c_2=\pi/2$ is a solution under the requirement
$\rp(\gamma_{\mathrm{B}}')=\ip(\gamma_{\mathrm{B}}')=0$. This is, as before,
a condition on the local operation $\op k_\mathrm{B}$. The other choice, $c_3
= \pi/2$, yields two possible solutions for $c_1$ and $c_2$, which again are
constrained by the local transformation $\op k_\mathrm{B}$. If
$\ip(\gamma_{\mathrm{B}}')^2=\frac{\mathcal{P}_{\mathrm{B}}(0)}{2}-\frac{1}{4}$,
the equations are solved with $c_1 = \pi/2$ and $c_2 = 0$, while for
$\rp(\gamma_{\mathrm{B}}')^2=\frac{\mathcal{P}_{\mathrm{B}}(0)}{2}-\frac{1}{4}$,
the solution reads $c_1 = 0$ and $c_2 = \pi/2$. This leaves us with three
possible control strategies, but all of them obey $c_1 + c_2 + c_3 = \pi$ and
therefore lead to a reset time of $T_\mathrm{min} = \pi/(2J)$. Finally we also
have to consider $\rp(\gamma_{\mathrm{B}}')^2\sin^2(c_2)
+ \ip(\gamma_{\mathrm{B}}')^2\sin^2(c_1)=0$, which is, in addition to $\sin(2
c_3) = 0$, the second solution to Eq.~\eqref{eq:third}. The only relevant case
corresponds to $\rp(\gamma_{\mathrm{B}}')=\ip(\gamma_{\mathrm{B}}')=0$, which
leads to $c_1=c_2=\pi/2$ and we retrieve one of the previous solutions. The
same analysis can be carried out for the other two equations, leading to the
same solutions. This indeed shows that considering all possibilities, it is
not possible to solve Eqs.~\eqref{eqs} with $c_1 + c_2 + c_3 < \pi$ and
therefore completes the proof. \qed
\end{proof}

Theorem~\ref{theo2} tells us that the shortest possible time $T_{\mathrm{min}}$
to reset the qubit is
$$
  T_{\mathrm{min}}=\frac{\pi}{2J}\,.
$$
We show in the main text that this lower bound can be attained for several
choices of control and interaction Hamiltonians, highlighting the tightness of
the bound. Attaining the bound for qubit reset is possible without the need for
local control over the ancilla, i.e., without complete controllability. Note
that the control procedure derived in the main text, i.e., the resonant
protocol, is completely different from the time-optimal strategy predicted by
Theorem~\ref{theo}, consisting in a concatenation of local hard pulses and free
evolutions~\cite{khaneja:2001}.

Most importantly, our results imply that local control over the qubit
ancilla does not allow to improve the reset time.

\subsection{Maximum purity and minimum reset time for qubit and ancilla on resonance}
\label{s:app:ana}
We explain now how to obtain the closed-form expression for the time evolution
of the qubit purity, Eq.~(\axref{8}) of the main text, given that the joint
qubit-ancilla dynamics is described by Hamiltonian~(\axref{6}). We present
a detailed derivation for $\op{H}_{\mathrm{int}} = J \left(\op{\sigma}_{1}
\otimes \op{\sigma}_{1}\right)$ and $\op{H}_{\mathrm{c}} = \pulse
\op{\sigma}_{1}$, i.e., $\OS = \OB = \Oc = \op{\sigma}_{1}$ and explain how to
extend the derivation to all other possible combinations of $\OS, \OB, \Oc \in
\{\op{\sigma}_{1}, \op{\sigma}_{2}, \op{\sigma}_{3}\}$ which fulfill
$\mathrm{dim}\{\mathfrak{a}\} = 2$, cf.\ Table~\ref{s:tab:dynLie}, presenting
all possible realizations for Hamiltonian~(\axref{6}).

We start by applying a transformation $\op{T} = \op{T}_{\mathrm{S}} \otimes
\uop_\mathrm{B}$, where $\op{T}_{\mathrm{S}}$ is chosen such that it
diagonalizes $\op{H}_{\mathrm{S}}$. The transformed Hamiltonian $\op{H}'
= \op{T}^{\dagger} \op{H} \op{T}$ for a constant and resonant field $\pulse$
becomes
\begin{align} \label{s:eq:H'}
  \op{H}'
  =
  \begin{pmatrix}
    \omega_{\mathrm{B}} & B & 0 & A
    \\
    B^{*} & 0 & A & 0
    \\
    0 & A & 0 & -B
    \\
    A & 0 & -B^{*} & -\omega_{\mathrm{B}}
  \end{pmatrix}
\end{align}
with $A = J \omega_{\mathrm{S}}/\omega_{\mathrm{B}}$ and $B
= 2 J \pulse/\omega_{\mathrm{B}}$. The resonance condition for this choice of
Hamiltonian implies $\pulse(t) = \pulse=\sqrt{\omega_\mathrm{B}^2
- \omega_\mathrm{S}^2}/2$ and, as a consequence, $|A|^2 + |B|^2 = J^2$. For the
constant resonant field, the time-evolution operator $\op{U}(t)$ can be
calculated analytically,
\begin{align} \label{s:eq:U}
  \op{U}(t)
  =
  e^{- \im \op{H}' t}
  =
  \begin{pmatrix}
    u_{11} & u_{12} & u_{13}     & u_{14}     \\
    u_{12} & u_{22} & u_{23}     & u_{13}     \\
    u_{13} & u_{23} & u_{22}^{*} & u_{12}^{*} \\
    u_{14} & u_{13} & u_{12}^{*} & u_{11}^{*}
  \end{pmatrix}
\end{align}
with
\begin{subequations} \label{s:eq:u_ij}
  \begin{align}\label{s:eq:u_11}
    u_{11}
    &=
    \frac{1}{2} \Bigg[
        \delta_{+} \cos(\Phi_{+})
      + \delta_{-} \cos(\Phi_{-})
      \notag \\
      & \quad \quad
      - \im \frac{1}{\eta_{+}} \left(
          \delta_{+} \omega_{\mathrm{B}}
          + \frac{2 |B|^{2}}{\Omega}
        \right) \sin(\Phi_{+})
      \notag \\
      & \quad \quad
      - \im \frac{1}{\eta_{-}} \left(
          \delta_{-} \omega_{\mathrm{B}}
          - \frac{2 |B|^{2}}{\Omega}
        \right) \sin(\Phi_{-})
    \Bigg],
  \end{align}
  \begin{align}
    u_{12}
    &=
    \frac{B}{\Omega} \big(\cos(\Phi_{+}) - \cos(\Phi_{-})\big)
    \notag \\
    & \quad
    -
    \im \frac{B}{2} \left(
        \frac{\delta_{+}}{\eta_{+}} \sin(\Phi_{+})
      + \frac{\delta_{-}}{\eta_{-}} \sin(\Phi_{-})
    \right),
  \end{align}
  \begin{align}
    u_{13}
    &=
    - \im \frac{A B}{\Omega}
    \left(
        \frac{1}{\eta_{+}} \sin(\Phi_{+})
      - \frac{1}{\eta_{-}} \sin(\Phi_{-})
    \right),
  \end{align}
  \begin{align}
    u_{14}
    &=
    - \im \frac{A}{2}
    \left(
        \frac{\delta_{+}}{\eta_{+}} \sin(\Phi_{+})
      + \frac{\delta_{-}}{\eta_{-}} \sin(\Phi_{-})
    \right),
  \end{align}
  \begin{align}
    u_{22}
    &=
    \frac{1}{2} \Bigg[
        \delta_{+} \cos(\Phi_{-})
      + \delta_{-} \cos(\Phi_{+})
      \notag \\
      & \quad \quad
      - \im \frac{2 |B|^{2}}{\eta_{+} \Omega} \sin(\Phi_{+})
      + \im \frac{2 |B|^{2}}{\eta_{-} \Omega} \sin(\Phi_{-})
    \Bigg],
  \end{align}
  \begin{align}
    u_{23}
    &=
    - \im \frac{A}{2}
    \left(
        \frac{\delta_{+}}{\eta_{-}} \sin(\Phi_{-})
      + \frac{\delta_{-}}{\eta_{+}} \sin(\Phi_{+})
    \right),
  \end{align}
\end{subequations}
where $\delta_{\pm} = 1 \pm \omega_{\mathrm{B}}/\Omega$, $\Omega
= \sqrt{\omega_{\mathrm{B}}^{2} + 4 |B|^2}$,
\begin{align} \label{s:eq:eta_pm}
  \eta_{\pm}
  &=
  \sqrt{%
    J^{2} + \frac{\omega_{\mathrm{B}}}{2}
    \left(%
      \omega_{\mathrm{B}} \pm \Omega
    \right)
  }
\end{align}
and $\Phi_{\pm}(t) = \eta_{\pm} t$. Note that $\Phi_{\pm}(t)$ is the only
time-dependent quantity in Eq.~\eqref{s:eq:u_ij}.

We can approximate Eqs.~\eqref{s:eq:U} and \eqref{s:eq:u_ij}, which are exact,
to derive an expression for the qubit purity,
\begin{equation}
  \label{eq:pur}
  \mathcal{P}_{\mathrm{S}}(t)
  =
  \tr\left\{%
    \tr_{\mathrm{B}}^{2}\left\{%
      \op{U}(t) \rZ \op{U}^{\dagger}(t)
    \right\}
  \right\}\,.
\end{equation}
Each element of the time-evolution operator is given by a sum of trigonometric
functions. We thus compare their amplitudes to identify the dominating terms. As
an illustration, the approximations will be explicitly shown for the amplitude
of the final term of $u_{11}$ in Eq.~\eqref{s:eq:u_11}, but the procedure is
equivalent for all other contributions. Using the relation $J = \sqrt{|A|^2
+ |B|^2}$, we express all variables in terms of $A$, $B$, and
$\omega_\mathrm{B}$. For the final term in Eq.~\eqref{s:eq:u_11}, this results
in
\begin{subequations}
  \begin{align}
    \delta_- = 1 - \frac{\omega_\mathrm{B}}{2\sqrt{\omega_\mathrm{B}^2 +
    |B|^2}},
  \end{align}
and
  \begin{align}
    \eta_- = \sqrt{|A|^2 + |B|^2 +
      \frac{\omega_\mathrm{B}^2}{2} - \frac{\omega_\mathrm{B}}{2}
    \sqrt{\omega_\mathrm{B}^2 + 4 |B|^2}}.
  \end{align}
\end{subequations}
Typically $J \ll \omega_\mathrm{B}$. Since $|B| \leq J$, this suggests an
expansion of all variables in $B$,
\begin{subequations}
  \begin{eqnarray}
    \delta_- &\approx& \frac{2 |B|^2}{\omega_\mathrm{B}^2} + \mathcal{O}(|B|^4),\\
\label{s:eq:approx_eta}
    \eta_- &\approx& |A| + \mathcal{O}(|B|^4),\\
    \Omega &\approx& \omega_\mathrm{B} + \frac{2|B|^2}{\omega_\mathrm{B}^2} +
    \mathcal{O}(|B|^4).
  \end{eqnarray}
\end{subequations}
With the approximated variables, we find
\begin{align}
  \frac{\im}{\eta_-}\left(\delta_- \omega_\mathrm{B} - 2|B|^2/\Omega\right)
  \approx \mathcal{O}(|B|^4)\,,
\end{align}
i.e., we can neglect the final term in Eq.~\eqref{s:eq:u_11}. Carrying out
similar approximations for the other amplitudes in Eq.~\eqref{s:eq:u_ij} leads
to
\begin{subequations} \label{s:eq:u_ij_approx}
  \begin{align}
    u_{11} &\approx \cos(\Phi_{+}) - \im \sin(\Phi_{+}),
    \\
    u_{12} &\approx 0,
    \\
    u_{13} &\approx 0,
    \\
    u_{14} &\approx 0,
    \\
    u_{22} &\approx \cos(\Phi_{-}),
    \\
    u_{23} &\approx \im \sin(\Phi_{-}).
  \end{align}
\end{subequations}
The corresponding approximated time-evolution operator allows us to
obtain a closed-form expression for the time evolution of the qubit purity
$\mathcal{P}_{\mathrm{S}}(t)$. In order to derive it, we additionally assume the
initial state of qubit and ancilla to be separable and the ancilla to be in
thermal equilibrium with its bath, cf. Eq.~(\axref{7}) in the main text. The
qubit purity, cf. Eq.~\eqref{eq:pur}, is then given by
\begin{align} \label{s:eq:PS_app}
  \mathcal{P}_{\mathrm{S}}(t)
  &=
  \left[%
    p_{\mathrm{S}}^{\mathrm{e}} p_{\mathrm{B}}^{\mathrm{e}} |u_{11}|^{2}
    +
    p_{\mathrm{S}}^{\mathrm{e}} p_{\mathrm{B}}^{\mathrm{g}} |u_{22}|^{2}
    +
    p_{\mathrm{S}}^{\mathrm{g}} p_{\mathrm{B}}^{\mathrm{e}} |u_{23}|^{2}
  \right]^{2}
  \notag \\
  &\quad
  +
  \left[%
    p_{\mathrm{S}}^{\mathrm{g}} p_{\mathrm{B}}^{\mathrm{g}} |u_{11}|^{2}
    +
    p_{\mathrm{S}}^{\mathrm{g}} p_{\mathrm{B}}^{\mathrm{e}} |u_{22}|^{2}
    +
    p_{\mathrm{S}}^{\mathrm{e}} p_{\mathrm{B}}^{\mathrm{g}} |u_{23}|^{2}
  \right]^{2}
  \notag \\
  &\quad
  +
  2 |\gamma_{\mathrm{S}}|^{2} |u_{11}|^{2} |u_{22}|^{2}.
\end{align}
Note that the qubit purity $\mathcal{P}_{\mathrm{S}}(t)$ depends only on
$\eta_{-}$. While $u_{11}$ depends on $\eta_{+}$, cf.
Eq.~\eqref{s:eq:u_ij_approx}, it enters $\mathcal{P}_{\mathrm{S}}(t)$ as
$|u_{11}|^{2} \approx 1$ such that the dependence on $\eta_+$ disappears when
inserting Eq.~\eqref{s:eq:u_ij_approx} and $\Phi_\pm=\eta_\pm t$ into
Eq.~\eqref{s:eq:PS_app}. Relabeling $\eta_-$ by $\eta$, we obtain
Eq.~(\axref{8}) of the main text.

\begin{table}
  \centering
  \caption{%
    Summary of the parameters $A$ and $B$ for all interactions
    $\op{H}_{\mathrm{int}} = J \left(\OS \otimes \OB\right)$ and local
    controls $\op{H}_{\mathrm{c}}(t) = \pulse(t) \Oc$ with $\OS, \OB, \Oc \in
    \{\op{\sigma}_{1}, \op{\sigma}_{2}, \op{\sigma}_{3}\}$. The third column
    indicates the form $(i)$ of the Hamiltonian $\op{H}_{i}'$, cf.
    Eq.~\eqref{s:eq:Hi'}. The last column states the minimal time
    $T_{\mathrm{min}}$, cf.\ Eq.~\eqref{s:eq:Tmin}, for purification of the
    qubit, evaluated with the same parameters as in Fig.~\axref{2}.
  }
  \begin{tabular*}{\linewidth}{c @{\extracolsep{\fill}} ccccc}
    \hline
    $\OS \otimes \OB$ & $\Oc$ & form & $A$ & $B$ & $T_{\mathrm{min}}$ \\
    \hline
    $\xx$ & $\op{\sigma}_{1}$ & (1) & $J\omega_{\mathrm{S}}/\omega_{\mathrm{B}}$     & $2 J\pulse/\omega_{\mathrm{B}}$                 & $46.9$ \\
    $\xx$ & $\op{\sigma}_{2}$ & (3) & $-\im J$                                       & $0$                                             & $15.7$ \\
    $\xx$ & $\op{\sigma}_{3}$ & (1) & $J$                                            & $0$                                             & $15.7$ \\
    \hline
    $\xy$ & $\op{\sigma}_{1}$ & (2) & $\im
    J\omega_{\mathrm{S}}/\omega_{\mathrm{B}}$ & $-2 \im J\pulse/\omega_{\mathrm{B}}$ & $46.9$ \\
    $\xy$ & $\op{\sigma}_{2}$ & (4) & $J$                                            & $0$                                             & $15.7$ \\
    $\xy$ & $\op{\sigma}_{3}$ & (2) & $\im J$                                        & $0$                                             & $15.7$ \\
    \hline
    $\xz$ & $\op{\sigma}_{1}$ & -   & -                                              & -                                               & - \\
    $\xz$ & $\op{\sigma}_{2}$ & -   & -                                              & -                                               & - \\
    $\xz$ & $\op{\sigma}_{3}$ & -   & -                                              & -                                               & - \\
    \hline
    $\yx$ & $\op{\sigma}_{1}$ & (3) & $\im J$                                        & $0$                                             & $15.7$ \\
    $\yx$ & $\op{\sigma}_{2}$ & (1) & $J\omega_{\mathrm{S}}/\omega_{\mathrm{B}}$ &   $2 J\pulse/\omega_{\mathrm{B}}$                   & $46.9$ \\
    $\yx$ & $\op{\sigma}_{3}$ & (3) & $\im J$                                        & $0$                                             & $15.7$ \\
    \hline
    $\yy$ & $\op{\sigma}_{1}$ & (4) & $-J$                                           & $0$                                             & $15.7$ \\
    $\yy$ & $\op{\sigma}_{2}$ & (2) & $\im J\omega_{\mathrm{S}}/\omega_{\mathrm{B}}$ & $-2\im J\pulse/\omega_{\mathrm{B}}$             & $46.9$ \\
    $\yy$ & $\op{\sigma}_{3}$ & (4) & $-J$                                           & $0$                                             & $15.7$ \\
    \hline
    $\yz$ & $\op{\sigma}_{1}$ & -   & -                                              & -                                               & - \\
    $\yz$ & $\op{\sigma}_{2}$ & -   & -                                              & -                                               & - \\
    $\yz$ & $\op{\sigma}_{3}$ & -   & -                                              & -                                               & - \\
    \hline
    $\zx$ & $\op{\sigma}_{1}$ & (1) & $-2 J\pulse/\omega_{\mathrm{B}}$               & $J\omega_{\mathrm{S}}/\omega_{\mathrm{B}}$      & $16.7$ \\
    $\zx$ & $\op{\sigma}_{2}$ & (1) & $-2 J\pulse/\omega_{\mathrm{B}}$               & $J\omega_{\mathrm{S}}/\omega_{\mathrm{B}}$      & $16.7$ \\
    $\zx$ & $\op{\sigma}_{3}$ & -   & -                                              & -                                               & -      \\
    \hline
    $\zy$ & $\op{\sigma}_{1}$ & (2) & $-2\im J\pulse/\omega_{\mathrm{B}}$            & $-\im J\omega_{\mathrm{S}}/\omega_{\mathrm{B}}$ & $16.7$ \\
    $\zy$ & $\op{\sigma}_{2}$ & (2) & $-2\im J\pulse/\omega_{\mathrm{B}}$            & $-\im J\omega_{\mathrm{S}}/\omega_{\mathrm{B}}$ & $16.7$ \\
    $\zy$ & $\op{\sigma}_{3}$ & -   & -                                              & -                                               & -      \\
    \hline
    $\zz$ & $\op{\sigma}_{1}$ & -   & -                                              & -                                               & - \\
    $\zz$ & $\op{\sigma}_{2}$ & -   & -                                              & -                                               & - \\
    $\zz$ & $\op{\sigma}_{3}$ & -   & -                                              & -                                               & - \\
    \hline
  \end{tabular*}
  \label{s:tab:Tmin}
\end{table}

For all the other combinations in Table~\ref{s:tab:dynLie} with
$\mathrm{dim}\{\mathfrak{a}\} = 2$, one just has to consider slightly modified
forms of the Hamiltonian $\op{H}_{1}' = \op{H}'$ in Eq.~\eqref{s:eq:H'}, namely
\begin{subequations} \label{s:eq:Hi'}
  \begin{align}
    \op{H}_{2}'
    =
    \begin{pmatrix}
      \omega_\mathrm{B} & B & 0 & A^*
      \\
      B^{*} & 0 & A & 0
      \\
      0 & A^* & 0 & -B
      \\
      A & 0 & -B^{*} & - \omega_{\mathrm{B}}
    \end{pmatrix},
  \end{align}
  or
  \begin{align}
    \op{H}_{3}'
    =
    \begin{pmatrix}
     \omega_{\mathrm{B}} & B & 0 & A^*
      \\
      B^{*} & 0 & A^* & 0
      \\
      0 & A & 0 & -B
      \\
      A & 0 & -B^{*} & -\omega_{\mathrm{B}}
    \end{pmatrix},
  \end{align}
  or
  \begin{align}
    \op{H}_{4}'
    =
    \begin{pmatrix}
      \omega_{\mathrm{B}} & B & 0 & A
      \\
      B^{*} & 0 & -A & 0
      \\
      0 & -A & 0 & -B
      \\
      A & 0 & -B^{*} & -\omega_{\mathrm{B}}
    \end{pmatrix},
  \end{align}
\end{subequations}
with $A$ and $B$ given in Table~\ref{s:tab:Tmin}.
The respective time-evolution operators are found to be
\begin{subequations}\label{s:eq:Ui}
  \begin{align}
    \op{U}_{2}(t)
    =
    \begin{pmatrix}
      u_{11} & u_{12} & u_{13}     &-u_{14}     \\
     -u_{12} & u_{22} & u_{23}     & u_{13}     \\
      u_{13} &-u_{23} & u_{22}^{*} &-u_{12}^{*} \\
      u_{14} & u_{13} & u_{12}^{*} & u_{11}^{*}
    \end{pmatrix},
  \end{align}
  or
  \begin{align}
    \op{U}_{3}(t)
    =
    \begin{pmatrix}
      u_{11} & u_{12} & u_{13}     &-u_{14}     \\
     -u_{12} & u_{22} &-u_{23}     & u_{13}     \\
      u_{13} & u_{23} & u_{22}^{*} &-u_{12}^{*} \\
      u_{14} & u_{13} & u_{12}^{*} & u_{11}^{*}
    \end{pmatrix},
  \end{align}
  or
  \begin{align}
    \op{U}_{4}(t)
    =
    \begin{pmatrix}
      u_{11} & u_{12} & u_{13}     & u_{14}     \\
     -u_{12} & u_{22} &-u_{23}     & u_{13}     \\
      u_{13} &-u_{23} & u_{22}^{*} & u_{12}^{*} \\
      u_{14} & u_{13} & u_{12}^{*} & u_{11}^{*}
    \end{pmatrix}\,,
  \end{align}
\end{subequations}
where the $u_{ij}$ refer to those used for $\op{U}_{1}(t) = \op{U}(t)$ in
Eq.~\eqref{s:eq:u_ij}, but need to be evaluated for the proper values of $A$ and
$B$, as listed in Table~\ref{s:tab:Tmin}. It is straightforward to check that
the qubit purity $\mathcal{P}_{\mathrm{S}}(t)$ for $\op{U}_{2}(t)$,
$\op{U}_{3}(t)$, and $\op{U}_{4}(t)$ is also given by Eq.~\eqref{s:eq:PS_app},
with the $u_{ij}$ modified as just described.

In order to determine the minimum time for purification, $T_{\mathrm{min}}$, we
demand $\dot{\mathcal{P}}_{\mathrm{S}}(t) = 0$ and
$\ddot{\mathcal{P}}_{\mathrm{S}}(t) < 0$. Inserting the $u_{ij}$ from
Eq.~\eqref{s:eq:u_ij_approx} into Eq.~\eqref{s:eq:PS_app}, we find
\begin{align} \label{s:eq:Tmin}
  T_{\mathrm{min}}
  =
  \frac{\pi}{2\eta_{-}} \approx \frac{\pi}{2|A|},
\end{align}
where, in the second step, we have used the approximation of
Eq.~\eqref{s:eq:approx_eta}. Equation~\eqref{s:eq:Tmin} implies that minimizing
the purification time corresponds to maximizing the amplitude of the
anti-diagonal of $\op{H}'$ in Eq.~\eqref{s:eq:H'} for the case $\OS = \OB = \Oc
= \op{\sigma}_{1}$. For the other cases, the formula for the minimal time
$T_{\mathrm{min}}$, cf.\ Eq.~\eqref{s:eq:Tmin}, is identical since
$\mathcal{P}_{\mathrm{S}}(t)$ depends only on the moduli of the $u_{ij}$.

Note that $T_{\mathrm{min}}$ is only determined by $\eta_{-}$, cf.\
Eq.~\eqref{s:eq:approx_eta}, and thus essentially by $|A|$. The latter should be
as large as possible in order for $T_{\mathrm{min}}$ to be minimal. Given
a specific choice of $\OS, \OB, \Oc \in \{\op{\sigma}_{1}, \op{\sigma}_{2},
\op{\sigma}_{3}\}$, $A$ and $B$ are determined by the resonance condition
$\lambda_{1} - \lambda_{0} = \omega_{\mathrm{B}}$, with $\lambda_{0}
<\lambda_{1}$ the eigenvalues of the qubit Hamiltonian $\op{H}_{\mathrm{S}}$,
and cannot be chosen at will. Rather, there exist certain combinations of
qubit-ancilla interaction and qubit control that maximize $|A|$. This is
summarized in Table~\ref{s:tab:Tmin}, which also indicates $B$ and the
respective form of the Hamiltonian, $\op{H}_{i}'$, $i=1,\dots,4$, cf.
Eqs.~\eqref{s:eq:H'} and~\eqref{s:eq:Hi'}, that generates the dynamics, cf.
Eqs.~\eqref{s:eq:U} and~\eqref{s:eq:Ui}.

The fact that, for all choices of qubit-ancilla interaction and qubit control,
only $|A|$ affects the minimal purification time $T_{\mathrm{min}}$ readily
explains the analytical and numerical results obtained in Fig.~\axref{2} of
the main text. Table~\ref{s:tab:Tmin} also shows that, while purification is
possible with several different interactions $\op{H}_{\mathrm{int}}
= J \left(\OS \otimes \OB\right)$, the specific choice of local control
$\op{H}_{\mathrm{c}}(t) = \pulse(t) \Oc$ crucially determines the achievable
purification time for that interaction.

\subsection{Superpositions of Pauli operators in $\op{H}_{\mathrm{int}}$ and $\op{H}_{\mathrm{c}}(t)$}
\label{s:app:superpos}
Next, as a generalization, we drop the restriction to Pauli operators ($\OS,
\OB, \Oc \in \{\op{\sigma}_{1}, \op{\sigma}_{2}, \op{\sigma}_{3}\}$) and instead
allow for operators of the form $\op{H}_{\mathrm{int}}
= J \left(\OS(\varphi_\mathrm{S}, \theta_\mathrm{S}) \otimes
\OB(\varphi_\mathrm{B}, \theta_\mathrm{B})\right)$ and a local control given by
$\op{H}_{\mathrm{c}}(t) = \pulse(t) \Oc(\varphi_\mathrm{c}, \theta_\mathrm{c})$
with
\begin{align}
  \label{eq:superpos}
  \op{O}_k (\varphi_{k}, \theta_{k})
  &=
  \cos(\varphi_{k}) \sin(\theta_{k}) \op{\sigma}_{1}
  +
  \sin(\varphi_{k}) \sin(\theta_{k}) \op{\sigma}_{2}
  \notag \\
  &\quad
  +
  \cos(\theta_{k}) \op{\sigma}_3\,,
\end{align}
where the angles are chosen as $\varphi_k \in [0, 2\pi]$ and $\theta_{k} \in [0,
\pi]$ with $k \in \{\mathrm{c},\mathrm{S},\mathrm{B}\}$. Performing the same
transformations as described above, i.e., diagonalizing the qubit Hamiltonian
$\op{H}_{\mathrm{S}}$, one arrives at
\begin{align} \label{s:eq:H'_gen}
  \op{H}'
  =
  \begin{pmatrix}
    \omega_{\mathrm{B}} - B_\mathrm{c}^+ & B_\mathrm{s}^+ e^{-\im
    \varphi_\mathrm{B}} & A_\mathrm{c}^{*} & A_\mathrm{s}^{*} e^{-\im \varphi_\mathrm{B}}
    \\
    B_\mathrm{s}^+ e^{\im \varphi_\mathrm{B}} & B_\mathrm{c}^+ &
    A_\mathrm{s}^{*} e^{\im \varphi_\mathrm{B}} & -A_\mathrm{c}^{*}
    \\
    A_\mathrm{c} & A_\mathrm{s} e^{-\im \varphi_\mathrm{B}} & -B_\mathrm{c}^- & B_\mathrm{s}^- e^{-\im \varphi_\mathrm{B}}
    \\
    A_\mathrm{s} e^{\im \varphi_\mathrm{B}} & -A_\mathrm{c} & B_\mathrm{s}^-
    e^{\im \varphi_\mathrm{B}}& -\omega_{\mathrm{B}} + B_\mathrm{c}^-
  \end{pmatrix},
\end{align}
where
\begin{subequations} \label{s:eq:AcAsB}
  \begin{align}
    A_{\mathrm{c}}
    &=
    \bar{A} \cos(\theta_\mathrm{B}),
    \\ \label{s:eq:As}
    A_{\mathrm{s}}
    &=
    \bar{A} \sin(\theta_\mathrm{B}),
    \\ \label{s:eq:A}
    \bar{A}
    &= J \bigg(
      \frac{\omega_+ + \omega_-}{2 \omega_\mathrm{B}}\sin(\theta_\mathrm{S}) \cos(\varphi_\mathrm{c}-\varphi_\mathrm{S})\nonumber\\
    &\phantom{=J \bigg(}
      -\frac{2 \pulse \cos(\theta_\mathrm{S})\sin(\theta_\mathrm{c})}{\omega_\mathrm{B}}\nonumber\\
    &\phantom{=J \bigg(}
      -\im \sin(\theta_\mathrm{S})\sin(\varphi_\mathrm{c}-\varphi_\mathrm{S})\bigg),
    \\
    B_\mathrm{c}^\pm
    &=
    J
    \frac{\cos(\theta_\mathrm{B})}{\Gamma_\pm^+}\left(\cos(\theta_\mathrm{S})\Gamma_\pm^- - \xi
    \omega_\pm \right),
    \\
    B_\mathrm{s}^\pm
    &=
    J
    \frac{\sin(\theta_\mathrm{B})}{\Gamma_\pm^+}\left(\cos(\theta_\mathrm{S})\Gamma_\pm^- + \xi
    \omega_\pm \right),
    \\
    \xi
    &=
    4 \pulse \sin(\theta_\mathrm{S}) \sin(\theta_\mathrm{c})
    \cos(\varphi_\mathrm{c} - \varphi_\mathrm{S}),
    \\
    \omega_\pm
    &=
    2 \pulse \cos(\theta_\mathrm{c}) + \omega_\mathrm{S} \pm
    \omega_\mathrm{B},
    \\
    \Gamma_\pm^+
    &=
    4 \pulse^2 \sin^2(\theta_\mathrm{c}) + \omega_\pm^2,
    \\
    \Gamma_\pm^-
    &=
    4 \pulse^2 \sin^2(\theta_\mathrm{c}) - \omega_\pm^2.
  \end{align}
\end{subequations}
Equation~ \eqref{s:eq:H'_gen} is a generalization of the Hamiltonians in
Eqs.~\eqref{s:eq:H'} and~\eqref{s:eq:Hi'}, and Eq.~\eqref{s:eq:AcAsB}
generalizes the variables accordingly. In particular, $\bar{A}$ is related to
$A$, which determines the minimum reset time for Pauli operators and is
listed in Table~\ref{s:tab:Tmin}, by $A=\bar{A} e^{\im \varphi_\mathrm{B}}
\sin(\theta_\mathrm{B})$.

\begin{figure}[tb]
  \centering
  \includegraphics{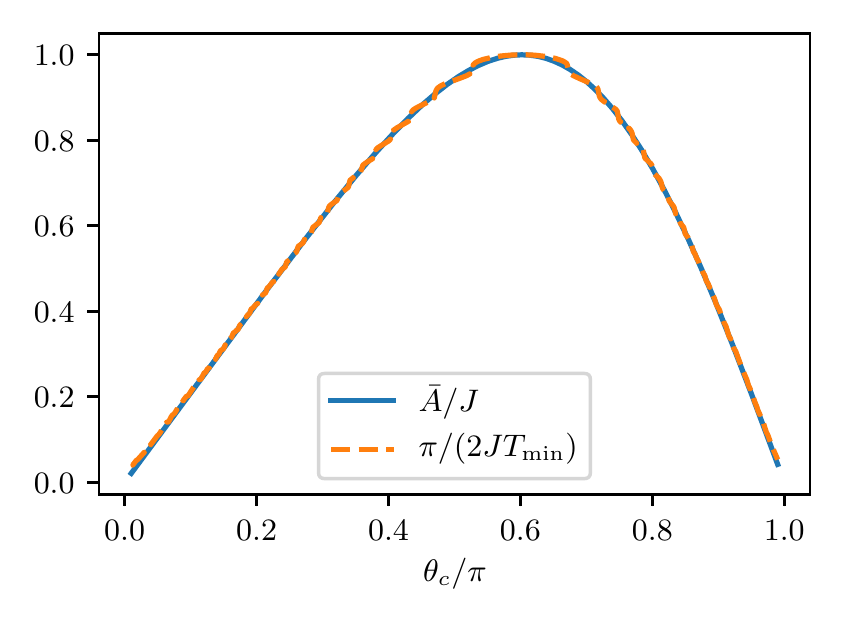}
  \caption{Behavior of $\bar{A}$ and the
    numerically obtained inverse purification time $1/T_\mathrm{min}$ as the
    control operator is varied via $\theta_c$ from a pure $\sigma_3$-control at
    $\theta_c=0$ to $\sigma_1$-control at $\theta_c = \pi/2$ back to a pure
    $\sigma_3$-control at $\theta_c=\pi$, but with opposite sign.
    The other angles are chosen such
    that there is a $\sigma_3\otimes\sigma_1$-interaction between qubit and
    ancilla. }
  \label{s:fig:A-Tmin}
\end{figure}

In the general case, Eq.~\eqref{eq:superpos}, it is not obvious how to obtain an
analytical expression for the time-evolution operator. Nevertheless,
generalizing the specific cases presented in Table~\ref{s:tab:Tmin} provides
some insight: According to Table~\ref{s:tab:Tmin}, purification requires
$\OB\neq\sigma_3$; otherwise the Cartan subalgebra is one-dimensional and purity
exchange between qubit and ancilla not possible. We therefore expect that
a $\sigma_3$-component in $\OB$ will not be helpful for faster purification. In
order to have no $\sigma_3$-component in $\OB$, we have to choose
$\theta_\mathrm{B} = \pi/2$. This implies in particular that the variables
$A_\mathrm{c}$ and $B_\mathrm{c}^\pm$ vanish, cf. Eq.~\eqref{s:eq:AcAsB}, which
eliminates the additional entries of the generalized Hamiltonian $\op{H}'$
compared to the special cases in Eqs.~\eqref{s:eq:H'} and~\eqref{s:eq:Hi'}.
Furthermore, this choice of $\theta_\mathrm{B}$ maximizes the magnitude of the
anti-diagonal, $A_\mathrm{s} = \bar{A}$. If we assume that, as before, an
optimal purification strategy is to maximize the magnitude of the anti-diagonal,
i.e., $|\bar{A}|$, also with respect to the other angles, then the angle
$\varphi_{\mathrm{B}}$ cannot be important, since it does not modify the
magnitude of any term in Eq.~\eqref{s:eq:H'_gen}, in other words,
$\varphi_\mathrm{B}$ is only responsible for a complex phase. Hence the task of
minimizing the purification time $T_\mathrm{min}$ in the generalized case
reduces to solving a three-angle problem involving $\theta_\mathrm{c},
\theta_\mathrm{S}, \varphi_\mathrm{c} - \varphi_\mathrm{S}$. Based on this
picture, we now conjecture that our result for the case of the Pauli operators
($\OS, \OB, \Oc \in \{\op{\sigma}_{1}, \op{\sigma}_{2}, \op{\sigma}_{3}\}$) ---
namely that maximal $|A|$, cf. Table~\eqref{s:tab:Tmin}, yields an optimal
solution --- holds true also for generalized interactions and control operators.
This is supported by numerical data, cf. Fig.~\ref{s:fig:A-Tmin}, showing that
maximal $1/T_\mathrm{min}$, hence minimal $T_\mathrm{min}$, concurs with maximal
$\bar{A}$ also for generalized control fields.

Given the importance of $\bar{A}$, respectively $A$, the lack of a physical
interpretation of this quantity is dissatisfying. In order to gain more insight,
we revisit Table~\ref{s:tab:Tmin}, which reveals maximal $A$ as a condition for
minimal purification time. Beyond that, given a certain interaction, the
purification time is also minimized if the commutator of the control
Hamiltonian, $\Oc\otimes\uop_\mathrm{B}$, and the interaction Hamiltonian,
$\op{H}_\mathrm{int} = J (\OS \otimes \OB)$, has maximum norm. The latter
information can be compressed into the quantity $C
=\frac{1}{2\sqrt{2}}\lVert[\OS, \Oc]\rVert$, which allows for a physical
interpretation: One can show that the norm of the commutator of $\Oc$ and
$\OS$ sets an upper limit to the energy (or, more specifically, heat) exchange
between the qubit and ancilla,
\begin{align}\label{s:eq:energy_exchange}
  |\dot{Q}|
  &=
  \tr\left( \dot{\op{\rho}}_\mathrm{S} \op{H}_\mathrm{S}
  \right)
  \nonumber\\
  &\leq \sqrt{2} J \left( \pulse \lVert\left[\OS,
  \Oc\right]\rVert + \frac{\omega_\mathrm{S}}{2} \lVert \left[\op{\sigma}_3,
  \OS\right] \rVert \right)\,.
\end{align}
One may now wonder whether $C$, relevant for the rate of heat exchange, is
related to $\bar{A}$, whose maximum is a condition for minimal purification
time.

\begin{figure*}[tb]
  \centering
  \includegraphics{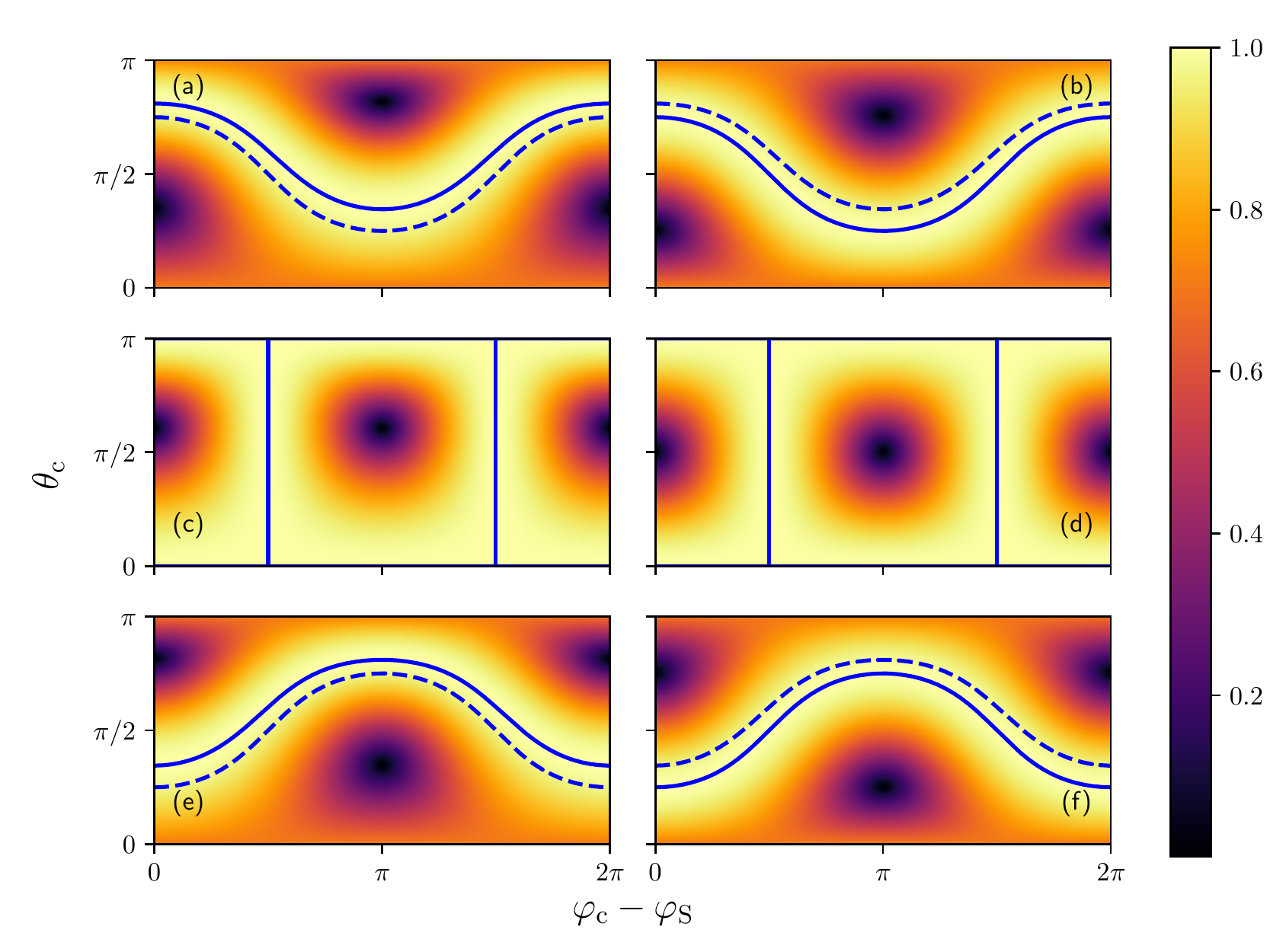}
  \caption{%
    The figure shows the values of $|\bar{A}|$ (left column) and
    $C$ (right column) for different interactions and control fields. These are
    determined by $\theta_\mathrm{S}=\pi/4$ in (a, b), $\theta_\mathrm{S}=\pi/2$
    in (c, d) and $\theta_\mathrm{S}=3\pi/2$ in (e, f) and the axes represent
    the angles $\theta_\mathrm{c}$ and $\varphi_\mathrm{c}-\varphi_\mathrm{S}$.
    The solid blue line indicates the maximum value of $1$ for the respective
    quantity, $|\bar{A}|$ or $|C|$, while the dashed line depicts the maximum
    value for the other quantity ($|C|$ or $|\bar{A}|$) to ease comparison.
    Except for (c) and (d), the lines do not coincide and therefore the limit on
    the energy exchange between qubit and ancilla, which is determined by $C$,
    cf.\ Eq.~\eqref{s:eq:energy_exchange}, does not fully explain the physical
    role of $\bar{A}$.
  }
  \label{s:fig:heatmap}
\end{figure*}

To check whether maximizing $|\bar{A}|$ corresponds to the same purification
strategy as maximizing $C$, we evaluate the two quantities for different
interactions and control fields and depict the corresponding values in
Fig.~\ref{s:fig:heatmap}. The results show that although the two quantities
behave very similarly, they only coincide for certain cases, namely those
indicated by the vertical lines in Fig.~\ref{s:fig:heatmap}(c,d). Incidentally,
these are exactly the pure Pauli operator choices for which the global minimum
of the purification time can be realized, cf. $T_\mathrm{min}=\pi/(2J)$
corresponding to $T_\mathrm{min}^{(1)}$ in the main text and resulting in the
value of $15.7$ in Table~\ref{s:tab:Tmin}. In these cases, both $C$ and
$|\bar{A}|$ are maximal. This underpins our claim of having identified
the globally minimal purification time. On the other hand, the different
behavior of $C$ and $|\bar{A}|$ as a function of the angles in general indicates
that the heat exchange argument made in Eq.~\eqref{s:eq:energy_exchange} alone
cannot fully explain the significance of $\bar{A}$. This requires further
investigation which will be the subject of future research.

\section{Generalization to ancillas with Hilbert space dimension $d> 2$}
\label{s:app:qudits} In the following, we exchange the two-level ancilla with an
ancilla with Hilbert space dimension $d> 2$. This is physically well justified,
since most quantum systems that can act as ancilla intrinsically possess more
than two levels. Here, we explore whether additional levels, energetically
above the ground and first excited state, are potentially beneficial for fast
on-demand reset. We discuss both aspects of reset, the reset time in
Sec.~\ref{subsec:tmind3}, focussing on $d=3$, as well as the achievable purity
in Sec.~\ref{subsec:purityd}, where we determine the maximal achievable purity
for arbitrary $d$.

\subsection{Minimal reset time} 
\label{subsec:tmind3}
In order to investigate whether an increase of the Hilbert space dimension of
the ancilla allows for faster purification, we first choose the ancilla to be
a qutrit ($d=3$). In the following, we focus on a local $\op{\sigma}_3$-control
on the qubit and generalize the original $\op{\sigma}_1 \otimes \op{\sigma}_1$
qubit-ancilla-interaction, which for two-level ancillas yields the globally
minimal purification time $T_{\mathrm{min}} = \pi/(2J)$, to the qubit-qutrit
case. To this end, we write the interaction Hamiltonian as $\op{H}_\mathrm{int}
= J[\sigma_1 \otimes (\op{a} + \op{a}^\dagger)]$, where $\op{a}$ and
$\op{a}^{\dagger}$ are the truncated lowering and raising operators,
respectively. For a qutrit in the basis $\{\ket{2}, \ket{1}, \ket{0}\}$,
$\op{a}$ reads
\begin{align}
  \op{a}
  =
  \begin{pmatrix}
    0 & 0 & 0 \\
    \sqrt{2} & 0 & 0 \\
    0 & 1 & 0
  \end{pmatrix}.
\end{align}
and the ancilla (or bath) Hamiltonian is given by
\begin{align}
  \op{H}_\mathrm{B}
  =
  \begin{pmatrix}
    \omega_{\mathrm{B},2} & 0 & 0\\
    0 & 0 & 0\\
    0 & 0 & - \omega_{\mathrm{B},1}
  \end{pmatrix},
\end{align}
with $\omega_{\mathrm{B},1}$ and $\omega_{\mathrm{B},2}$ the transition
frequencies between the qutrit's ground and first excited state, respectively
first and second excited state. As before, we assume uncorrelated initial
thermal states on system and ancilla.

\begin{figure}[tb]
  \centering
  \includegraphics{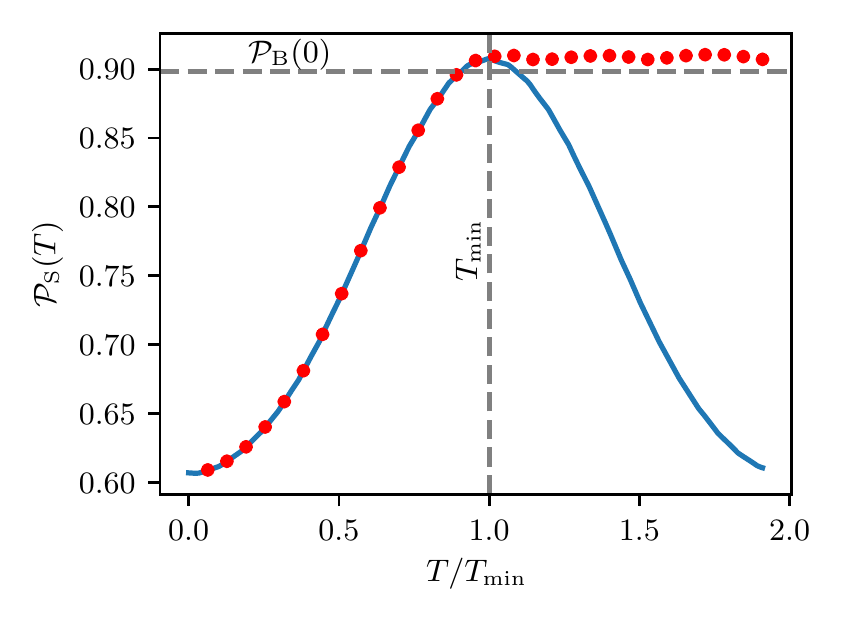}
  \caption{%
    Purity evolution of a qubit interacting with a qutrit ancilla under a
    constant, resonant field (blue line). The resonance condition is set with
    respect to the $\ket{0}\leftrightarrow\ket{1}$ transition in the qutrit, as
    these are the most populated levels. With numerically optimized fields we
    also obtained the maximal purity for specific times (red dots). Parameters
    as in Fig.~\axref{2} and $\omega_{\mathrm{B},1} = 3$ and
    $\omega_{\mathrm{B},2} = 2$.
  }
  \label{s:fig:Pmax-qutrit}
\end{figure}

In order to examine a possible change in the minimal reset time due to the
addition of a third level, we have numerically maximized the qubit purity for
different final times $T$, cf. Fig.~\ref{s:fig:Pmax-qutrit}. The highest purity
in Fig.~\ref{s:fig:Pmax-qutrit} is observed for times equal or larger than
$T_{\mathrm{min}}$, i.e., the minimum time to achieve maximum purity is
identical to the case of a two-level ancilla. Analogous simulations for $d=4$
(data not shown) yield the same minimum purification time. We therefore
conjecture that $T_{\mathrm{min}}$ is independent of the ancilla dimension.

Moreover, we observe that, for all $T < T_{\mathrm{min}}$ in
Fig.~\ref{s:fig:Pmax-qutrit}, a resonant guess field (blue line) yields the
maximally achievable qubit purity, where the resonance is taken with respect to
$\omega_{\mathrm{B},1}$. We therefore conjecture, based on numerical evidence,
that resonant fields remain an optimal reset strategy also for ancillas with
Hilbert space dimension larger than two.

Figure~\ref{s:fig:Pmax-qutrit} allows for another important observation: While
the minimum reset time $T_{\mathrm{min}}$ is unchanged for $d=3$ compared to
$d=2$, the maximal achievable purity increases, cf.\ the red dots and dashed
horizontal line which corresponds to the maximal qubit purity in case of
a two-level ancilla. Remarkably, the maximum qubit purity is no longer upper
bounded by the initial ancilla purity, $\mathcal{P}_{\mathrm{B}}(0)$. This
raises the question whether the maximal qubit purity can be further enhanced by
allowing for even larger $d$. We answer this question in the affirmative in the
following section by examining the connection between maximal achievable system
purity and ancilla Hilbert space dimension.

\subsection{Maximally achievable qubit purity employing qudit ancillas}
\label{subsec:purityd}
As before, we assume an initially separable state of qubit and ancilla, $\rSB
= \rS \otimes \rB$, where the ancilla Hilbert space has dimension
$d_\mathrm{B}$. We now seek to find those unitary transformations $\op{U} \in
\mathrm{SU}(2d_\mathrm{B})$ such that the purity of the time-evolved qubit,
\begin{align}
  \rS'
  =
  \tr_{\mathrm{B}} \left\{\op{U} \rSB \op{U}^{\dagger}\right\}
  =
  \dmap_{\rB} [\rS]\,,
\end{align}
is maximized --- irrespective of the Hamiltonian and control fields that
generate those unitaries. First, we demand the state $\rSB'$ yielding maximal
purity to be separable, $\rSB' = \rS' \otimes \rB'$. Since the ultimate goal
is to purify the qubit, which includes erasing all correlations with the
ancilla, it is natural to consider a separable target state.

\begin{figure*}[tb]
  \centering
  \includegraphics{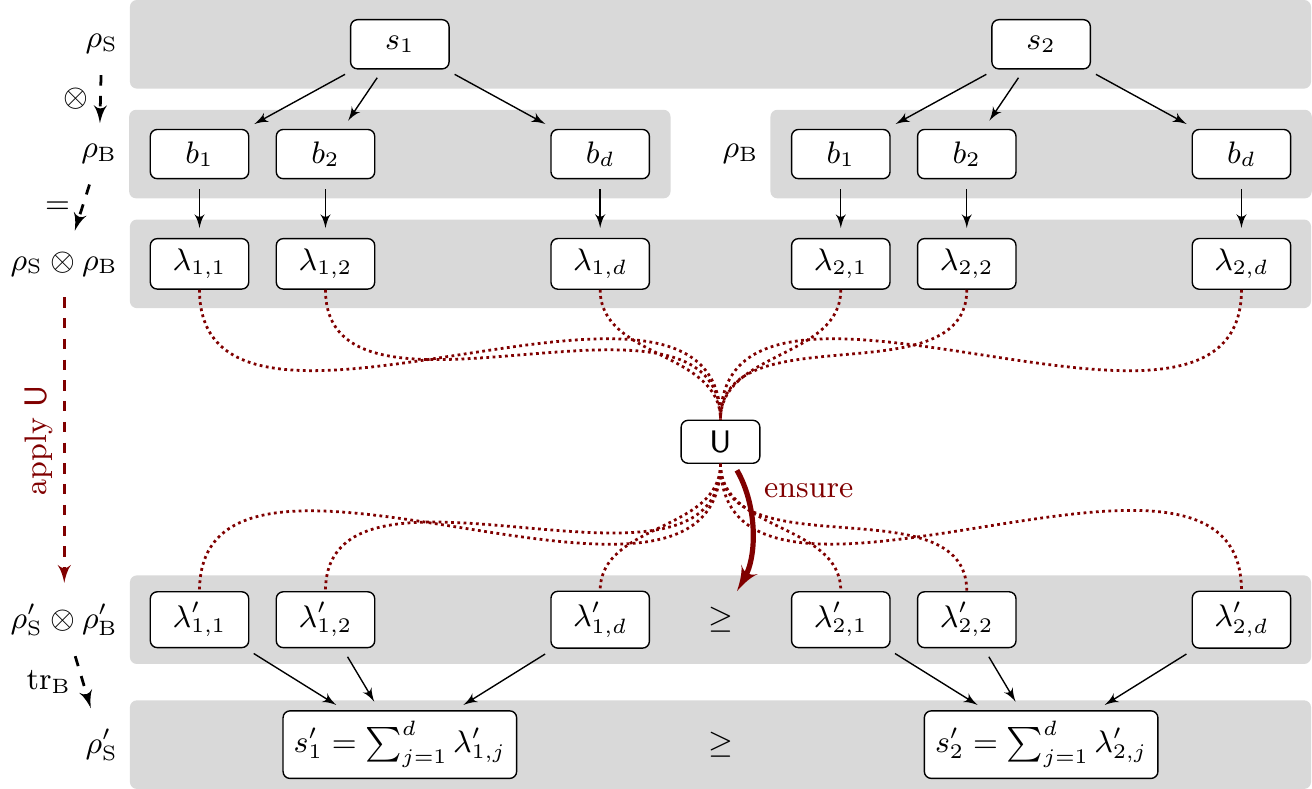}
  \caption{%
    Sketch of the reshuffling operation on the joint qubit ancilla spectrum that
    maximizes the qubit purity.
  }
  \label{s:fig:reshuffle}
\end{figure*}

Next, we write the initial states of qubit and ancilla in the respective
eigenbases, $\{\ket{s_{1}}, \ket{s_{2}}\}$ and $\{\ket{b_{1}}, \dots,
\ket{b_{d}}\}$,
\begin{align}
  \rS
  =
  \sum_{i=1}^{2} s_{i} \ket{s_{i}} \bra{s_{i}},
  \qquad
  \rB
  =
  \sum_{j=1}^{d_\mathrm{B}} b_{j} \ket{b_{j}} \bra{b_{j}}\,.
\end{align}
The eigenvalues of $\rS$ and $\rB$ obey $0 \leq s_{1}, s_{2} \leq 1$ and $0 \leq
b_{1}, \dots, b_{d} \leq 1$ with $s_{1} + s_{2} = 1$ and
$\sum_{j=1}^{d_\mathrm{B}} b_{j} = 1$. Thus, the joint initial state reads
\begin{align} \label{s:eq:rSB_in}
  \rSB
  =
  \sum_{i=1}^{2} \sum_{j=1}^{d_\mathrm{B}}
  \lambda_{i,j} \ket{s_{i}} \bra{s_{i}} \otimes \ket{b_{j}} \bra{b_{j}},
  \quad
  \lambda_{i,j}
  =
  s_{i} b_{j}.
\end{align}
Since we demand the final state to be separable as well, we can write, using
prime for all transformed quantities,
\begin{align} \label{s:eq:rSB_out}
  \rSB'
  =
  \sum_{i=1}^{2} \sum_{j=1}^{d_\mathrm{B}}
  \lambda_{i,j}' \ket{s_{i}'} \bra{s_{i}'} \otimes \ket{b_{j}'} \bra{b_{j}'},
\end{align}
Due to the unitary nature of the transformation, the spectra $\{\lambda_{i,j}\}$
and $\{\lambda_{i,j}'\}$ of the joint states $\rSB$ and $\rSB'$ are identical.
However, $\op{U}$ allows for spectral reordering, which is what ultimately
allows purification of the qubit. Due to the separability of $\rSB'$, the qubit
purity becomes
\begin{align} \label{s:eq:PS}
  \mathcal{P}_{\mathrm{S}}
  =
  \tr\left\{\rS'^{2}\right\}
  =
  \sum_{i=1}^{2} s_{i}'^{2},
  \qquad
  s_{i}'
  =
  \sum_{j=1}^{d_\mathrm{B}} \lambda_{i,j}'.
\end{align}
We can interpret the elements of the qubit spectrum after the reset, $\{s_{1}',
s_{2}'\}$, as entries of a vector, $\myvec{s'} = (s_{1}', s_{2}')^{\top}$. By
means of Karamata's inequality and using the concept of majorization, we can
construct an operation $\op{U}$ that yields maximal $\mathcal{P}_{\mathrm{S}}$.
\begin{mydef*}[Majorization] 
  Let $\myvec{a}, \myvec{b} \in \mathbb{R}^{d}$, $d \in \mathbb{N}$, and let
  $\myvec{a}^{\downarrow}, \myvec{b}^{\downarrow}$ be reshuffled vectors
  $\myvec{a}, \myvec{b}$ with their elements sorted in descending order. The
  vector $\myvec{a}$ majorizes the vector $\myvec{b}$, i.e., $\myvec{a} \succ
  \myvec{b}$, if $\sum_{n=1}^{d} a_{n} = \sum_{n=1}^{d} b_{n}$ and
  $\sum_{n=1}^{k} a_{n}^{\downarrow} \geq \sum_{n=1}^{k} b_{n}^{\downarrow}$ for
  all $k=1,\dots,d$.
\end{mydef*}
\begin{theorem}[Karamata's inequality] 
  Let $I$ be an interval of the real line, $I \subset \mathbb{R}$, and $f:
  I \rightarrow \mathbb{R}$ a convex function. If $\myvec{a} \succ \myvec{b}$
  for $\myvec{a}, \myvec{b} \in I^{d}$, $d \in \mathbb{N}$, then
  $$\sum_{n=1}^{d} f(a_{i}) \geq \sum_{n=1}^{d} f(b_{i})\,.$$ If $f$ is strictly
  convex then $\sum_{n=1}^{d} f(a_{i}) = \sum_{n=1}^{d} f(b_{i})$ iff $\myvec{a}
  = \myvec{b}$.
\end{theorem}
Taking into account the strict convexity of the function $f(x) = x^{2}$, we know
that, if $\myvec{s'}^{*} \succ \myvec{s'}$, then
\begin{align}
  \mathcal{P}_{\mathrm{S}}(\myvec{s'}^{*})
  =
  \sum_{i} f(s_{i}'^{*})
  \geq
  \sum_{i} f(s_{i}')
  =
  \mathcal{P}_{\mathrm{S}}(\myvec{s'})\,.
\end{align}
Thus, for maximal qubit purification, we have to find a vector $\myvec{s'}^{*}$
that majorizes all other vectors accessible via spectral reshuffling from
$\{\lambda_{i,j}\}$ to $\{\lambda_{i,j}'\}$. With the constraint $s_{1}'^{*}+
s_{2}'^{*} = 1$ for a qubit, majorization of $\myvec{s'}^{*}$ implies, without
loss of generality, maximization of $s_{1}'^{*} = \sum_{j=1}^{d_\mathrm{B}}
\lambda_{1,j}'$, cf. Eq.~\eqref{s:eq:PS}. Hence, the unitary transformation
$\op{U} \in \mathrm{SU}(2d_\mathrm{B})$ needs to reshuffle the $2d_\mathrm{B}$
elements from the initial set $\{\lambda_{1,1}, \dots, \lambda_{1,d_\mathrm{B}},
\lambda_{2,1}, \dots, \lambda_{2,d_\mathrm{B}}\}$, which can be in any order,
such that the first half of the reordered set $\{\lambda_{1,1}', \dots,
\lambda_{1,d_\mathrm{B}}', \lambda_{2,1}', \dots, \lambda_{2,d_\mathrm{B}}'\}$
contains the $d_\mathrm{B}$ largest eigenvalues and the second half the
remaining ones. The reshuffling is sketched in Fig.~\ref{s:fig:reshuffle}.
Karamata's inequality then guarantees that this gives the largest achievable
qubit purity.

A possible choice for $\op{U}$ is given by the permutation matrix that
transforms the vector $\myvec{\lambda} = (\lambda_{1,1}, \dots,
\lambda_{1,d_\mathrm{B}}, \lambda_{2,1}, \dots,
\lambda_{2,d_\mathrm{B}})^{\top}$ into the vector $\myvec{\lambda'}
= (\lambda_{1,1}', \dots, \lambda_{1,d_\mathrm{B}}', \lambda_{2,1}', \dots,
\lambda_{2,d_\mathrm{B}}')^{\top}$. Note that $\op{U}$ is not unique since, for
instance, any local operation on either the qubit or the ancilla leaves the
qubit purity invariant.

The protocol described so far can easily be generalized to arbitrary qudit-qudit
systems. One only needs to ensure that the unitary $\op{U}$ reshuffles all
elements from $\{\lambda_{i,j}\}$ such that the vector $\myvec{s'}^{*}
= (s_{1}'^{*}, s_{2}'^{*}, \dots)^{\top}$ majorizes all other accessible
vectors.

The above results have an interesting implication. Let us assume that at least
half (rounded up) of the initial eigenvalues $\{b_{1}, \dots,
b_{d_\mathrm{B}}\}$ of $\rB$ are zero. As a consequence, at least half of the
eigenvalues $\{\lambda_{i,j}\}$ of the initial qubit-ancilla state will be zero.
Assuming full unitary controllability, these eigenvalues can be reshuffled such
that the first half, namely $\lambda_{1,j}'$ with $j=1,\dots,d_\mathrm{B}$,
contains all non-zero eigenvalues whereas the other half, $\lambda_{2,j}'$ with
$j=1,\dots,d_\mathrm{B}$, contains only zeros. Equation~\eqref{s:eq:PS} then
implies $s_{1}' = 1$ and $s_{2}' = 0$ , i.e., we obtain a pure state
$\mathcal{P}_{\mathrm{S}} = 1$ --- independently of the initial qubit state
$\rS$ and despite the fact that $\rB$ is mixed. For ancillas with Hilbert space
dimension larger than two, it is therefore possible to purify the qubit beyond
a simple swap of purities.

We can generalize our observation to arbitrary Hilbert space dimensions of
system and ancilla. The following proposition reveals the relation between
Hilbert space dimensions and achievable purity.
\begin{myprop*}
  Let $\hil_{\mathrm{S}}$ and $\hil_{\mathrm{B}}$ be Hilbert spaces with
  dimension $d_{\mathrm{S}}$ and $d_{\mathrm{B}}$, respectively, and
  $\liou_{\hil_{\mathrm{S}}}$ and $\liou_{\hil_{\mathrm{B}}}$ their
  corresponding Liouville spaces. Let $\rB \in \liou_{\hil_{\mathrm{B}}}$ be
  a density matrix with $\lceil d_{\mathrm{B}} (d_{\mathrm{S}}-1)
  / d_{\mathrm{S}} \rceil$ eigenvalues below $\epsilon /(2 d_{\mathrm{B}}
  (d_{\mathrm{S}}-1))$ with small $\epsilon > 0$. Then, for all density matrices
  $\rS \in \liou_{\hil_{\mathrm{S}}}$, there exists a $\op{U} \in
  \mathrm{SU}(d_{\mathrm{S}} d_{\mathrm{B}})$ such that
  \begin{align}
    1 - \tr\left\{ \rS'^{2} \right\}
    \leq
    \epsilon,
    \qquad
    \rS'
    =
    \tr_{\mathrm{B}} \left\{%
      \op{U} (\rS \otimes \rB) \op{U}^{\dagger}
    \right\},
  \end{align}
  i.e., the purity of $\rS'$ gets $\epsilon$-close to unity.
\end{myprop*}
\begin{proof}
  Let $\{\lambda_{i,j}\}$ and $\{\lambda_{i,j}'\}$ be the spectra of the
  states $\rSB = \rS \otimes \rB$ and $\rSB' = \op{U} \rSB \op{U}^{\dagger}
  = \rS' \otimes \rB'$, defined in the same way as in Eqs.~\eqref{s:eq:rSB_in}
  and~\eqref{s:eq:rSB_out}. Since $\op{U}$ is unitary, these spectra are
  identical up to reshuffling. The purity of $\rS'$ reads
  $\mathcal{P}_{\mathrm{S}} = \tr\left\{ \rS'^{2} \right\}
  = \sum_{i=1}^{d_{\mathrm{S}}} s_{i}'^{2}$ with $s_{i}'
  = \sum_{j=1}^{d_{\mathrm{B}}} \lambda_{i,j}'$. Exploiting unit trace,
  $\sum_{i=1}^{d_{\mathrm{S}}} s_{i}' = 1$, we can rewrite the purity as
  $\mathcal{P}_{\mathrm{S}} = 1 + 2 \sum_{i=2}^{d_{\mathrm{S}}} s_{i}'
  (s_{i}'-1)$, from which we deduce that
  \begin{align}
    1 - \mathcal{P}_{\mathrm{S}}
    \leq
    \epsilon
    \quad \Leftrightarrow \quad
    \sum_{i=2}^{d_{\mathrm{S}}} s_{i}' (s_{i}'-1)
    \leq
    \frac{\epsilon}{2}.
  \end{align}
  Since $s_{i}' \in [0,1]$, $i=1,\dots,d_{\mathrm{S}}$, one option to ensure
  the latter inequality is
  \begin{align} \label{s:eq:approx}
    \sum_{i=2}^{d_{\mathrm{S}}} s_{i}'
    \leq
    \frac{\epsilon}{2},
  \end{align}
  since $\sum_{i=2}^{d_{\mathrm{S}}} s_{i}' (s_{i}'-1) \leq
  \sum_{i=2}^{d_{\mathrm{S}}} s_{i}'$. We now assume that $n < d_{\mathrm{B}}$
  eigenvalues of $\rB$ are below some small $\delta > 0$. Thus, we know that
  $d_{\mathrm{S}} n$ eigenvalues in $\{\lambda_{i,j}\}$ will be below $\delta$.
  If $d_{\mathrm{S}} n \geq d_{\mathrm{B}} (d_{\mathrm{S}}-1)$, which implies $n
  \geq d_{\mathrm{B}} (d_{\mathrm{S}}-1) / d_{\mathrm{S}}$, then we can
  reshuffle $\{\lambda_{i,j}\}$ into $\{\lambda_{i,j}'\}$ in such a way that all
  elements $\lambda_{i,j}' \leq \delta$ for $i=2,\dots,d_{\mathrm{S}}$ and
  $j=1,\dots,d_{\mathrm{B}}$. Hence, we find $s_{i}'
  = \sum_{j=1}^{d_{\mathrm{B}}} \lambda_{i,j}' \leq \delta d_{\mathrm{B}}$,
  $i=2,\dots,d_{\mathrm{S}}$. Plugging this into Eq.~\eqref{s:eq:approx}, we
  obtain
  \begin{align}
    \sum_{i=2}^{d_{\mathrm{S}}} s_{i}'
    \leq
    (d_{\mathrm{S}}-1) d_{\mathrm{B}} \delta
    \leq
    \frac{\epsilon}{2}.
  \end{align}
  This can be guaranteed if $\delta \leq \epsilon / (2 d_{\mathrm{B}}
  (d_{\mathrm{S}}-1))$ and the proposition follows. \qed
\end{proof}

We now apply the proposition to the purification of a qubit ($d_{\mathrm{S}}=2$)
by an ancilla with Hilbert space dimension $d_{\mathrm{B}}$. For the special
case of a two-level ancilla, $\lceil d_{\mathrm{B}} (d_{\mathrm{S}}-1)
/ d_{\mathrm{S}} \rceil = \lceil 2 (2-1) / 2 \rceil = 1$, i.e., the initial
ancilla state $\rB$ needs to have one small eigenvalue in order for the final
qubit state $\rS'$ to become almost pure. In other words, the reset error
becomes small only for initially almost pure ancillas. For $d_\mathrm{B}=3$, we
find $\lceil d_\mathrm{B}/2 \rceil = 2$, which means that we still require an
almost pure initial ancilla state for successful purification of the qubit.
Starting from $d_\mathrm{B} = 4$, it is still possible to successfully purify
even if more than one eigenvalue of $\rB$ is large. In other words for
$d_\mathrm{B} \geq 4$, the initial ancilla state $\rB$ does not have to be pure
initially in order to fully purify the qubit.

It should be noted, however, that for an initial thermal population
distribution on the ancilla, a qubit purity of exactly one can only be reached if
$\rho_\mathrm{B}$ is initially pure. This result agrees with the findings of
Ref.~\cite{Wu2013}, which show that if the system and a thermal bath are
initially factorized then cooling the system to a pure state is only possible if
the bath is initially at zero temperature. Our proposition allows us to go one
step further by showing that it is possible to get $\epsilon$-close to unit
purity, even if the bath is initially in a mixed state, under certain conditions
(namely ensuring that $\delta \leq \epsilon / (2 d_{\mathrm{B}}
(d_{\mathrm{S}}-1))$, where $\delta$ represents an upper bound for a sufficient
number of ancilla eigenvalues at initial time).

Moreover, our findings are in accordance with a theorem showing that cooling the
system $\epsilon$-close to a pure state is possible as long as the bath initial
state is sufficiently close to a so-called "subsystem pure state
initialization"~\cite{Ticozzi2014}. While Ref.~\cite{Ticozzi2014} quantifies
closeness primarily via the trace distance, our derivation yields a bound on the
achievable system purity as a function of the spectral properties of the initial
bath state.


%